# How effective is graphene nanopore geometry on DNA sequencing?

**Vahid Satarifard[1], Masumeh Foroutan[*1], Mohammad Reza Ejtehadi[2]**

[1] Department of Physical Chemistry, School of Chemistry, College of Science, University of Tehran, Tehran, Iran

[2] Faculty of Physics, Sharif University of Technology, Tehran, Iran

**Abstract**

In this paper we investigate the effects of graphene nanopore geometry on homopolymer ssDNA pulling process through nanopore using steered molecular dynamic (SMD) simulations. Different graphene nanopores are examined including axially symmetric and asymmetric monolayer graphene nanopores as well as five layer graphene polyhedral crystals (GPC). The pulling force profile, moving fashion of ssDNA, work done in irreversible DNA pulling and orientations of DNA bases near the nanopore are assessed. Simulation results demonstrate the strong effect of the pore shape as well as geometrical symmetry on free energy barrier, orientations and dynamic of DNA translocation through graphene nanopore. Our study proposes that the symmetric circular geometry of monolayer graphene nanopore with high pulling velocity can be used for DNA sequencing.

[*] Corresponding author, foroutan@ut.ac.ir (M. Foroutan), Telefax: +982166495291

# 1. Introduction

DNA sequencing by cheaper and faster techniques has an emerging role in therapeutics and personalized medicine. Furthermore it can help to improve our perception of inheritance, genetic risk factors, diseases and evolutionary biology [1-4]. Nanopore-based DNA sequencing is one of the controversial approaches and has different modalities for sequencing, like ionic current sensing, transverse electron tunneling, optical readout and capacitive sensing. Nanopore sensors for nucleic acid analysis are free label, free amplification and real-time [5-7].

Among the numerous host materials for fabricating nanopores, the graphene membrane is one of the best candidates. Graphene is a two-dimensional sheet of carbon atoms bonded by $sp^2$ hybrid orbitals [8]. Graphene membrane thickness is analogous to the distance between neighboring nucleotides in DNA i.e. 0.32–0.52 nm, and this makes the graphene as a perfect candidate for DNA sequencing.

To assess the merit of graphene apertures for DNA sequencing, several groups have reported experimental works [5], MD simulations [9, 10] as well as first-principle-based studies [11]. First experimental proofs of DNA sensing with graphene nanopore were reported in 2010 [12-15]. In the recent years, several MD studies were performed to reproduce the experimental observation. Recently the influence of some affective factors on DNA sequencing has been studied like applied voltage [9, 10, 16-18], DNA conformation [9], DNA homopolymer sequences [9, 10, 16, 17], pore charge [9], pore size [9, 10, 16, 18], pore shape [10], membrane thickness [10, 16, 17], salt concentration [16-18]. Until now, none of experimental works as well as simulation studies have not presented acceptable evidence of single nucleotide.

Recently, for homopolymer ssDNA pulling process through nanopore, the effects of harmonic constant and pulling velocity were studied on force profile in SMD simulation [19]. Their results show that guanine and adenine in the DNA strand can be identified and cytosine and thymine remain indistinguishable. Also, some first-principle-based studies were implemented for study of electronic properties [20-23] of graphene nanopore and transverse tunneling conductance [21, 24-26] for sequencing purpose. Their results demonstrate that nucleotides can be identified based on the specific transverse tunneling conductance of nucleotides.

Because of the tiny sub-nanometric size of pores, the symmetry and shape of nanopore can play an important role in DNA translocation. From the experimental point of view, sculpting process (fabricating nanopore) by electron beam or ion beam, can introduce defects and amorphization of graphene in insulator membrane; subsequently in many cases the final shape of pore is asymmetric [27-29]. To investigate the effects of graphene nanopore shape on detecting long and short DNA fragments, lately Freedman et al., studied the use of multilayer graphene including nanopores with GPC formed around the nanopore edge. They reported on the translocation of double stranded and single stranded DNA through such graphene pores and showed that the DNA single stranded translocates much slower than other allowing detection of extremely short fragments,25 nucleotides in length [30]. Their findings suggested that the kinetic and controllable properties of graphene nanopores under sculpting conditions can be used to further enhance the detection of DNA analytes. Very recently, Wang et al. (2014), used single Si defect in graphene lattice to catalyze the dissociation of carbon atoms from graphene [31] and made the arbitrarily shape of nanopore in graphene. Their research goal was to fabricate stable functional devices at the atomic scale precision which is an excellent candidate for molecular detection applications, such as rapid DNA sequencing.

Nevertheless, in DNA translocation through subnanometric graphene pore under electric field or pulling process, the effects of pore shape are not well understood and the recent researches are limited to comparing the ssDNA translocation through elliptical and circular pores [10] and the obtained results show that in the case of elliptical pores, a higher bias lead to slower translocation due to frustration, by trapping the DNA in a conformation unfavorable for translocation.

Also the letter [32] that has been published very recently, the effect of graphene nanopore geometry on DNA sequencing has been studied by SMD simulation. The authors have claimed that identification of nucleotides is possible via their SMD simulations. However, based on some experimental [33, 34] as well as simulation [19] study, these results are under some criticism that will be addressed in the section 3 of present paper.

In this study, we investigate the effects of axially symmetric and asymmetric pore shapes, as well as 5 layers graphene polyhedral crystal (5L-GPC) shape of nanopore on DNA translocation through subnanometric graphene pore by SMD simulations. The pulling force profile, moving fashion of ssDNA, work done in irreversible DNA pulling and orientations of DNA bases near the nanopore will be assessed to show that how effective is graphene nanopore geometry on DNA sequencing.

## 2. Simulation details and methods

All MD simulations were performed using the LAMMPS package [35] and structures were visualized using the VMD package [36] CHARMM27 [37] force field and TIP3P [38] model were used for modeling of DNA and water molecules, respectively. Carbon atoms of the graphene sheet treated as aromatic carbons (type CA) in the CHARMM27 and were fixed during simulations. The system was first minimized for 2000 steps, then was heated to 300 K and was equilibrated for 50 ps under NPT ensemble. After a NPT relaxation, 50 ps equilibration was performed in an NVT ensemble. The Nosé-Hoover barostat and thermostat were applied to maintain the pressure and temperature at 1 atm and 300 K, with damping coefficients 1 ps$^{-1}$ and 0.1 ps$^{-1}$, respectively. Time step of 1 fs was employed.

The Particle-Particle Particle -Mesh method, with the cutoff distance of 10 Å, was used to calculate the electrostatic interactions [39]. The characteristics of 22 systems that were studied are summarized in Table 1. The box size was $38 \times 38 \times 114$ Å$^3$ for simulation number 1-14, $38 \times 38 \times 218$ Å$^3$ for simulation number 15-18 and $38 \times 38 \times 183$ Å$^3$ for simulation number 19-22, Table 1.

Several SMD simulations were performed to exert a harmonic force on the z-axis of phosphorus atom as the SMD atom of the first nucleotide of the ssDNA homopolymer chains including six nucleotides namely, poly(dT)$_6$, poly(dC)$_6$, poly(dA)$_6$ and poly(dG)$_6$ homopolymers.

In each simulation, the ssDNA was pulled at a constant velocity of 10 Å/$ns$ as well as 5 Å/$ns$, and a spring constant of 1000 $pN$/Å. In this work, all atom MD simulations were performed to provide an atomic level description of DNA pulling through different shapes of graphene nanopore. Fig 1 shows different shapes of graphene aperture that were used in this study including monolayer graphene membrane and five layers graphene polyhedral crystal (5L-GPC) shape of nanopore.

As shown in Fig 1a, some carbon atoms of the single layer graphene arbitrarily removed to generate axially symmetric shape of nanopores 1 and 3 (with the D$_{2h}$ symmetry) as well as asymmetric shape of nanopore 2 (with the C$_{2h}$ symmetry. Fig 1b shows the cross section and top view of 5L-GPC. In order to study the effect of thickness and shape of nanopore in graphene membrane, five layer graphene polyhedral crystals (5L-GPC) was used. Pore area is 0.786 nm$^2$, 0.835 nm$^2$ and 0.933 nm$^2$ for pore 1, 2 and 3, respectively.

The pore 3 and 5L-GPC were fabricated by removing carbon atoms whose coordinates fulfilled the condition $\sqrt{x^2 + y^2} \leq r$ and $\sqrt{x^2 + y^2} \leq r + \tan(\theta) \times |z|$, circle and double-cone equation, respectively, where r is the diameter of the pore and $\theta$ is pore aperture angle. The pore geometry of central layer in 5L-GPC is similar to geometry of pore 3. By changing the angle of pore, one can adjust structure of double-cone in multilayer graphene. The pore 1 and 2 were constructed by removing carbon atoms to build the noted pore, without any specific equation. All nanopore fabricating procedure was conducted with VMD package [36]. Graphene membrane was located in the xy plane and in Cartesian coordinate origin (0, 0, 0).

## 3. Results

### 3.1 Time-position dependence of SMD atom in ssDNA

To understand the ssDNA moving fashion, time-Z position values of SMD atom, phosphorus atom of the first nucleotide, were recorded during the pulling procedure. As shown in Fig 2a-c, four ssDNA homopolymer, poly(dT)$_6$, poly(dC)$_6$, poly(dA)$_6$ and poly(dG)$_6$, pulled through monolayer graphene at a constant velocity of 10 Å/$ns$. In Fig 2e, the movement fashion of poly(dT)$_6$ through pore 1 at different constant velocities of 10 Å/$ns$ and 5 Å/$ns$, were illustrated. Also Fig 2f shows the movement fashion of poly(dA)$_6$ through 5L-GPC at a constant velocity of 10 Å/$ns$ and in different degrees of pore angles $\theta = 15°\ and\ \theta = 30°$.

### 3.2 Work done in irreversible DNA pulling procedure

To evaluate the work in irreversible DNA pulling through different shapes of graphene nanopores, the running integral over the pulling force in direction of the spring is recorded. The work calculated by numerical integration in every step, based on the following equation:

$$W = v \int_{t'}^{t''} dt f(t) \tag{1}$$

where $v$ is the pulling velocity.

The work done during the irreversible pulling simulations can be considered the overestimated attribute, as same trends of free energy barrier calculated from irreversible (nonequilibrium) simulations. Also using Jarzynski's equality [40], one can evaluate an approximate potential of mean force, by multiple trajectories, as a free energy barrier in SMD simulations. Reaction coordinate in Fig 3 indicates the position of SMD atom in pulling procedure.

Figs 3a-3d, illustrate work done in irreversible pulling of poly(dA)$_6$, poly(dG)$_6$, poly(dT)$_6$ and poly(dC)$_6$, respectively, through monolayer graphene at a constant velocity of 10 Å/$ns$. Fig 3e, compares work of irreversible pulling of poly(dA)$_6$ through monolayer pore 3 and multilayer 5L-GPC, at a constant velocity of 10 Å/$ns$.

### 3.3 Force profiles

To obtain sequence-specific recognition at single nucleotide resolution, the exerted force on spring was measured. The ssDNA homopolymers were pulled by the spring at constant velocity. Once a nucleotide reaches to entrance of the nanopore, because of the hydrophobic interaction between carbons of graphene and nucleotides, ssDNA traps in *cis* chamber of the nanopore. The *cis* chamber is defined as the initial position of DNA next to graphene.

In pulling procedure, once bases reach to pore entrance, the applied force on the spring increases approximately linear, to the extent that one of the nucleotides passes through the nanopore and enters to next chamber of nanopore, afterwards the force was dropped rapidly.

Figs 4a-4f illustrate the force profile of poly(dA)$_6$, poly(dG)$_6$, poly(dT)$_6$ and Poly(dC)$_6$ with pulling velocity of 10 Å/$ns$, for nanopores 1, 2 and 3, respectively. In the case of pores 1 and 2, the sharp force peaks are observed in a wide range of force values for all of the homopolymers; 6-10 nN for pore 1 and 4-8 nN for pore 2.

For nanopore 3 as the most symmetric and the biggest nanopore in this study, the pulling force profiles of homopolymers with narrow ranges of force values for purin and pyrimidine bases are obtained; 4-5 nN for purine [poly(dA)$_6$, poly(dG)$_6$] and 3-4 nN for pyrimidine [poly(dT)$_6$ and Poly(dC)$_6$].

Fig 4g illustrates the force profile of poly(dT)$_6$ pulling through pore 1 at low pulling velocity of 5 Å/ns. In this Fig the horizontal arrow indicates the translocations of two nucleotides which are released at the same time through nanopore. Fig 4h shows the force profile of poly(dA)$_6$ for 5L-GPC entrance with two pore angles, $\theta = 15°$ and $\theta = 30°$.

### 3.4 The orientations of DNA bases near the pore

In order to achieve detailed insight into the behavior of nucleotides near the pore and before translocation, a set of basis vectors is used in molecular plan, to subsequently define two angles, $\beta$ and $\gamma$, which specify the orientation of each nucleotides [10]. Fig 5a illustrates the set of basis vectors including $e_1$ and $e_2^*$. The vector $e_1$ is defined by three atoms coordinate and $e_2$ defines as $e_2 \equiv e_2^* - (e_2^*.e_1)e_1$. Atom coordinates is sampled from each 100 frames of entire trajectory. As shown in Fig 5b, to investigate of the tilt of bases, the $\beta$ is defined as the angle between the vectors $e_3$ and $z$, where $e_3$ defines as $e_3 \equiv e_1 \times e_2$ and $z$ is the unit vector direction normal to the graphene membrane. To describe the rotation of molecules plan, the $\gamma$ is defined as the angle between the vectors $e_2$ and $r$, where $r$ defined as $r \equiv (e_3 \times z) \times e_3$. Figs 6a-6h indicate the probability density distribution (PDD) of bases angles near the pore and before translocation. Bases are in a distance of 3 Å from graphene pore when DNA traps in cis side of nanopore. Fig 6a, 6c, 6e and 6g, illustrate the PDDs of $\gamma$, and Fig 6b, 6d, 6f and 6h, illustrate the PDDs of $\beta$ in the SMD simulation of poly(dA)$_6$, poly(dG)$_6$, poly(dT)$_6$ and Poly(dC)$_6$, respectively.

## 4. Discussion

### 4.1 Time-position dependence of SMD atom in ssDNA

Simulation results proposed a ratchet-like fashion in adopted parameters for pulling of all homopolymers. Similar base-by-base ratcheting has also been proposed for ssDNA inside solid-state nanopores [19, 32, 41]. This style of moving is favorable to control of nucleotides movement in the stepwise passing. Figs 2a -2f illustrate the effect of pore shape on ratchet-like motion of ssDNA. Obviously, the pore shape has high impact on varying the temporarily position and velocity of SMD atom.

Each skip in Z position of phosphorous atom indicates the translocation of one nucleotide at high pulling velocity 10 Å/$ns$. Whereas under slow pulling velocity, i.e. 5 Å/$ns$, some jumps indicate that two nucleotides translocates simultaneously. The vertical arrow in Fig 2e, shows the passage of two nucleotides which released simultaneously through nanopore.

At high pulling velocity, 10 Å/$ns$, the last nucleotide of all ssDNA chains passes without any signals. While, at low pulling velocity, 5 Å/$ns$, some nucleotides pass without any signals and it is due to favorable orientations in translocation of bases. This observation emphasizes on the importance of pulling velocity for DNA sequencing. At low pulling velocity, because of passing of several bases, the information of DNA sequence is missed.

In the case of 5L-GPC, the movement of ssDNA is highly influenced by the slop of double-cone entrance. In small pore angle i.e. $\theta = 15°$, the ten skips in the movement diagram are related to passing of six nucleotides. As Fig 2f shows, in pore angle $\theta = 30°$, the number of jumps in the movement diagram decreases to five.

Using of multilayer graphene with double cone shape decreases the sensitivity of DNA sequencing. These results for 5L-GPC are of interest for sequencing purpose, since these types of skips are undesirable.

In the case of pore 3, the time of DNA translocation through nanopore for all nucleotides is nearly the same, indicating the possibility of identification of homopolymers with this axially symmetric pore shape. Furthermore, the translocation of one nucleotide through pores 1, 2 and 5L with different times shows that these geometries are not applicable for sequencing purposes. This issue will extend by further calculations and will present in next sections.

**4.2 Work done in irreversible DNA pulling procedure**

One can see some skips in all work diagrams; each skip is related to rapidly passage of a nucleotide through nanopore and indicates the work done to release each nucleotide in a ratchet fashion. The local slopes of work diagrams in Fig 3 demonstrate the different conformations of nucleotides in translocation time. Moreover, work diagrams obviously show different free energy barriers for pulling of identical ssDNA through different apertures of monolayer graphene nanopores. Our results propose that tiny alteration in the shape of graphene nanopore, also increasing of thickness of graphene membrane will change the work done in irreversible DNA pulling

For pulling poly(dA)$_6$ through pores 1, 2, 3 and 5L-GPC, the differences in work diagram between the initial and final conformations are 4000 kcal/mol, 2800 kcal/mol, 1120 kcal/mol and 1360 kcal/mol, respectively as shown in Figs 3a and 3e. These work values reveal that, by changing of the ratio of longitudinal to transverse radius of nanopore as well as increment of the thickness of graphene membrane the work increases, therefore we expect the same trend for free energy barrier. For asymmetric pores and bigger pores like pore 3, the free energy barrier for ssDNA translocation decreases. These trends are seen for all other homopolymers.

**4.3 Force profiles**

The results propose that the axially symmetric rhombic pore with a big ratio of longitudinal to transverse radius (pore 1) and the asymmetric pore (pore 2) are not good choices for DNA sequencing. However the axially symmetric circular pore (pore 3) can be considered as an appropriate choice for identifying different nucleotides. These results illustrate the importance of effects of pulling velocity value and the geometry of nanopore on DNA sequencing.

In the case of 5L-GPC with $\theta = 15°$, since the number of force peaks are more than passed bases, 10 peaks for six nucleotides, by decreasing the entrance angle, the resolution of nanopore reduces. At $\theta = 30°$, five peaks are seen for six nucleotides, but none of them can identify the nucleotides of poly(dA)$_6$ homopolymer.

In Fig 4, each sharp decrease change in force indicates the passing of one nucleotide through nanopore, except at $\theta = 15°$ as Fig 4h shows. As the force profiles of homopolymers for pores 1 and 2 and 5L-GPC show the single base resolution peaks are appeared at different forces, and therefore distinguishing between nucleotides is not possible by using these nanopores.

As the above results show, pore 3 can identify the purine and pyrimidine bases completely. By repeating the pulling procedure several more times with different initial conditions, the purines bases can be identified but the pyrimidine bases remain indistinguishable [19, 32].

The results of force profile could be explained with available conformations of each nucleotide near the nanopore, which was investigated by assessing orientations as will be addressed later.

For investigation of sequence specific signal of nucleotides, the Fourier transform of the saw-tooth form force profiles were carried out for different pore shapes, the highest peaks in the spectrum just indicate to the length of simulation time, and the second and third one are related to next harmonics in force profiles. So there is no valuable information in Fourier transform of the force profile in such a tiny time scale simulations.

Recently, Zhang et al. (2014) have studied the effect of graphene nanopore geometry on DNA sequencing by considering DNA fragments including A, T, C, G and 5-methylcytosine (MC) using SMD simulation.

It seems that Zhang et al. (2014) have reported a case study, with specific sequence of ssDNA (CCMC MCCC TTAAG), as well as some parameters like the parameters for the spring.

In some cases, in the presence of purine bases (A and G) among the sequence, the error bars of measured force peaks changes. For example for the sequence ssDNA as CCT AAG TCG

ATG, that have been investigated by SMD [19], the existence of the first and the second G among the sequence, cause the big drop in force peak for further bases, T and A, respectively, which indicates the importance of the sequence of bases in DNA for nanopore sensing method. In addition, the speed of pulling in their SMD simulation [19, 32] seems to be too fast for a realistic atomic force microscopy experiment [34] and it is not well understood what is the role of pulling speed and consequence of reducing speed in results of SMD simulation.

Above cases can be considered as big challenges that have been addressed in real experiments for long sequence of DNA using this method. Therefore it seems that Zhang et al. (2014) have made a wrong claim about identification of nucleotides by SMD simulation.

**4.4 The orientations of DNA bases near the pore**

The PDDs of $\gamma$ angle, propose roughly the same distribution of angles for purine and pyrimidine bases. Purine bases prefer to rotate the molecular plan in most of the time $240° < \gamma < 280°$ and occasionally $70° < \gamma < 100°$, and the purine bases could not pass through other angles. In the case of pyrimidine bases, molecular plan prefer to rotate in most of the time $200° < \gamma < 300°$ and occasionally $70° < \gamma < 110°$, and the pyrimidine bases could not pass through other angles.

These results indicate a wide distribution *of* $\gamma$ angle in pyrimidine and purine bases. The wider distribution *of* $\gamma$ angle in pyrimidine bases can be assigned to its smaller molecular plan that causes the extra movement. The PDDs of $\gamma$ angle do not illustrate the difference between geometries for pulling of homopolymers and also demonstrate the tiny effect of rotation on conformational changes in DNA translocation through nanopores.

The obtained PDDs show the $\beta$ angle with different distributions for purine and pyrimidine bases, and also illustrate the profound dependency of these distributions on the geometry of nanopores.

All of the nucleotides could tilt their molecular plan at angles $90° < \gamma < 180°$, except for Adenine that can tilt its molecular plan at angles $0° < \gamma < 70°$.

In all PDDs of $\beta$ angle, one can see wider distribution of $\beta$ angle for pore 3 compared to pores 1 and 2. Pore 3 as an axially symmetric and the biggest nanopore, provides much chance for bases to pass through nanopore with different orientations.

Homopolymers can pass through nanopore 3 with different orientations, thus they have little free energy barrier during pulling procedure. Consequently, the force peaks for pulling homopolymers appears in the same ranges as shown in Figs 4e and 4f.

Furthermore, the PDDs of $\beta$ angles for pores 1 and 2 indicate that homopolymers pass through these pores with thin angle distributions. It means that ssDNA traps in back of the nanopore in an unfavorable conformation for passing. Therefore, because of the geometry restriction of nanopores 1 and 2, homopolymers encounter with higher free energy barrier and need to higher forces during pulling procedure.

The obtained results for orientations of the DNA bases near the pores illustrate the high correlation between the PDDs of molecular plan tilt, $\beta$ angle, and geometry of a pore.

Also, the wide PDD for pore 3 demonstrates the possibility of nucleotides identification using this kind of pore and also explains the obtained results for work and force profiles.

## 5. Conclusions

In summary, the effects of nanopore shape on the ssDNA pulling through graphene nanopores were investigated using SMD simulations. The results showed the high impact of symmetry and shape of nanopore on the pulling force profiles and also on the free energy barrier during pulling procedure. The obtained results for orientations of the ssDNA bases near the pores illustrated the large impact of tilt of molecular plan on conformational change during translocation.

The pulling force profiles indicated that, symmetric circular nanopore embedded in monolayer graphene at high pulling velocity can be used for characterizing and identifying bases. Moreover, by repeating pulling procedure and using force profiles, adenine and guanine can be identified but cytosine and thymine still remain indistinguishable. We expect that our results can motivate further experimental research using atomic force microscopy experiments and optical tweezers.

# Figs Captions

**Fig . 1** Different shapes of graphene aperture were used in this study. (a) Monolayer graphene membrane, (b) Nanopore with five layers graphene polyhedral crystal (5L-GPC)

**Fig . 2** Time-Zposition dependence of SMD atom (phosphorus atom of the first nucleotide) of (a) Poly(dA)6 (b) Poly(dG)6 (c) Poly(dT)6 (d) Poly(dC)6.(e) effect of pulling velocity on Poly(dT)$_6$ nucleotides jumping, the horizontal arrow was indicated passing of two nucleotides which was released at the same time through nanopore (f) effect of double-cone entrance slope on Poly(dA)6 movement

**Fig . 3** Work done in irreversible DNA pulling of (a) Poly(dA)6 (b) Poly(dG)6 (c) Poly(dT)6 (d) Poly(dC)$_6$. In diagram (e) effect of double-cone entrance on work done in pulling of Poly(dA)6

**Fig . 4** Pulling force profile of (a) Poly(dA)6 and Poly(dG)6 for Pore1 (b) Poly(dT)6 and Poly(dC)$_6$ for Pore1 (c) Poly(dA)6 and Poly(dG)6 for Pore2 (d) Poly(dT)6 and Poly(dC)6 for Pore2 (e) Poly(dA)6 and Poly(dG)6 for Pore3 (f) Poly(dT)6 and Poly(dC)6 for Pore 3 (g) Poly(dT)6 for Pore1 at low pulling velocity of 5 Å/ns, the horizontal arrow was indicated passing of two nucleotides which was released at the same time through nanopore (h) Poly(dA)6 for 5L-GPC entrance in different pore angle

**Fig . 5** To describe the orientations of DNA bases near the pore, (a) A set of basis vectors was defined by three atoms coordinate. (b)To investigate the tilt of bases, the β is defined as the angle between the vectors $e_3$ and z, where z is the unit vector direction normal to the graphene membrane; and to describe the rotation of molecules plan, the γ is defined as the angle between the vectors $e_2$ and r. Adapted with permission from (REFERENCE [10]). Copyright (2012) American Chemical Society

**Fig . 6** The orientations, PDD of tilt and rotation, of bases near the pore. The tilt and rotation of the bases were defined by the angles β and γ respectively; their explanations are illustrated in Fig 5
Fig 6a, 6c, 6e and 6g, illustrate the PDDs of *γ*, and 6b, 6d, 6f and 6h, illustrate the PDDs of β in SMD simulation of poly(dA)6, poly(dG)6, poly(dT)6 and Poly(dC)6

# Tables

**Table 1** Characteristics of the simulated systems

| Simulation number | Pore number | ssDNA type | Pulling velocity (Å/ns) | Number of graphene layers | Time(ns) |
|---|---|---|---|---|---|
| 1 | 1 | Poly(dA)6 | 10 | Monolayer | 6 |
| 2 | 1 | Poly(dG)6 | 10 | Monolayer | 6 |
| 3 | 1 | Poly(dT)6 | 10 | Monolayer | 6 |
| 4 | 1 | Poly(dC)6 | 10 | Monolayer | 6 |
| 5 | 2 | Poly(dA)6 | 10 | Monolayer | 6 |
| 6 | 2 | Poly(dG)6 | 10 | Monolayer | 6 |
| 7 | 2 | Poly(dT)6 | 10 | Monolayer | 6 |
| 8 | 2 | Poly(dC)6 | 10 | Monolayer | 6 |
| 9 | 3 | Poly(dA)6 | 10 | Monolayer | 6 |
| 10 | 3 | Poly(dG)6 | 10 | Monolayer | 6 |
| 11 | 3 | Poly(dT)6 | 10 | Monolayer | 6 |
| 12 | 3 | Poly(dC)6 | 10 | Monolayer | 6 |
| 13 | 1 | Poly(dT)6 | 5 | Monolayer | 11 |
| 14 | 2 | Poly(dC)6 | 5 | Monolayer | 11 |
| 15 | 2 | Poly(dA)12 | 10 | Monolayer | 9 |
| 16 | 3 | Poly(dA)12 | 10 | Monolayer | 9 |
| 17 | 2 | Poly(dG)12 | 10 | Monolayer | 9 |
| 18 | 3 | Poly(dG)12 | 10 | Monolayer | 9 |
| 19 | 5L-GPC | Poly(dA)6 | 10 | Five layers | 5 |
| 20 | 5L-GPC | Poly(dG)6 | 10 | Five layers | 5 |
| 21 | 5L-GPC | Poly(dT)6 | 10 | Five layers | 5 |
| 22 | 5L-GPC | Poly(dC)6 | 10 | Five layers | 5 |

**Fig 1**

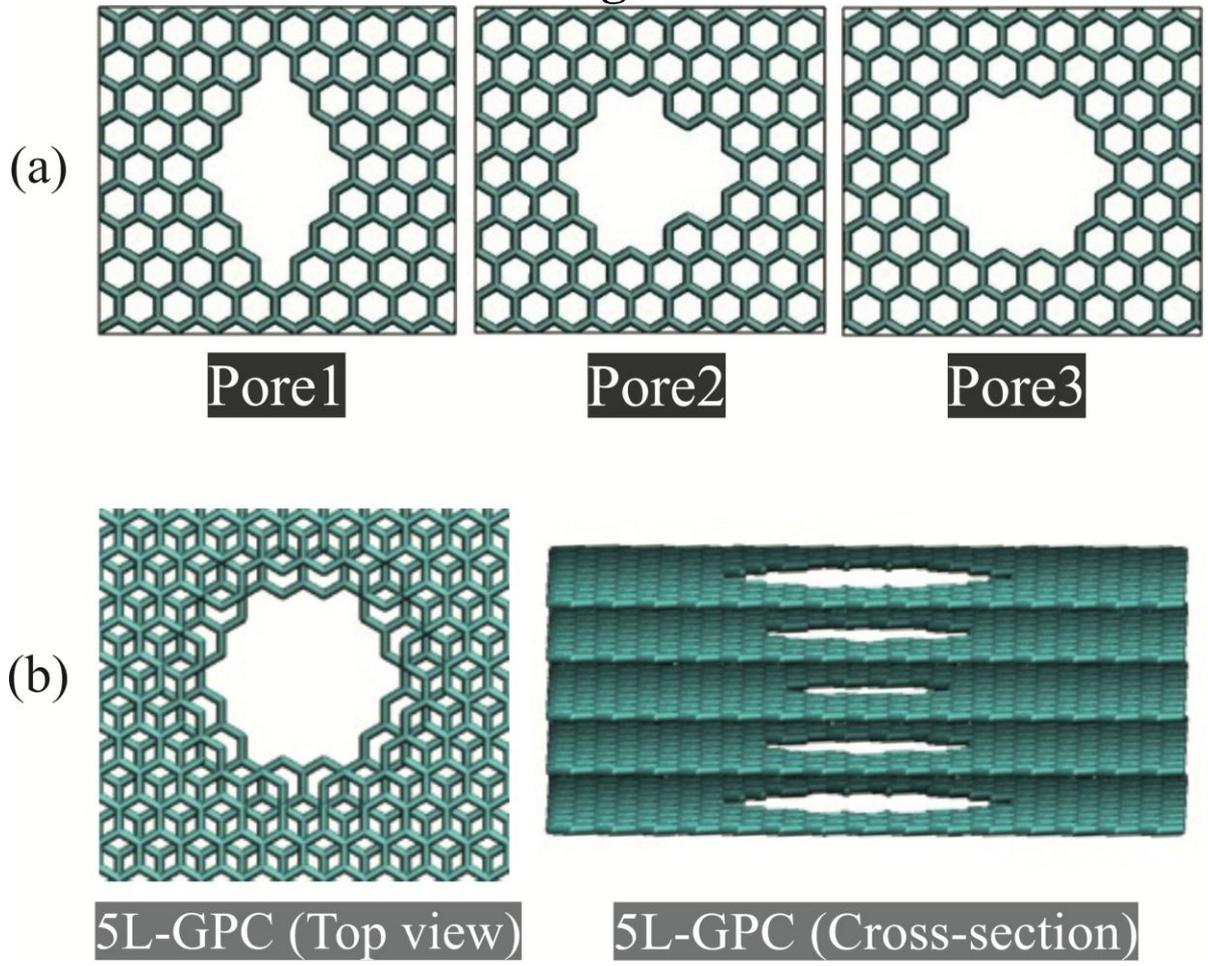

**Fig 2a**

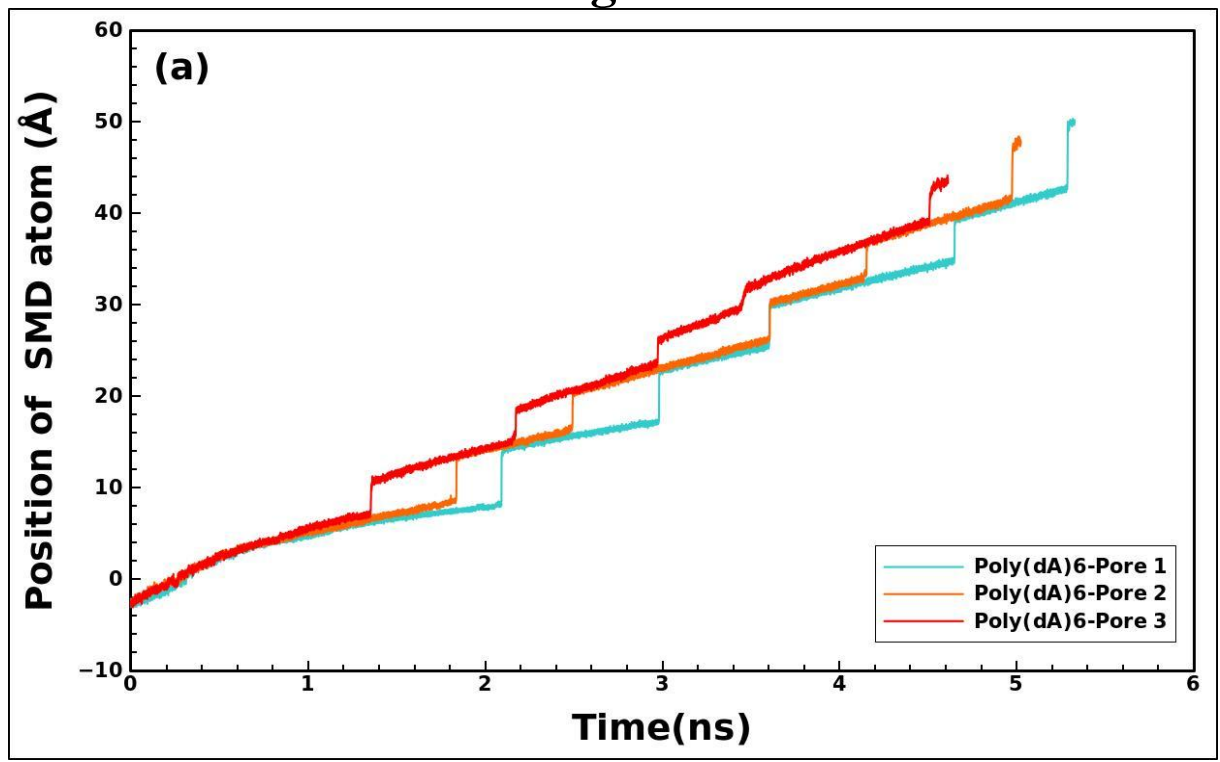

**Fig 2b**

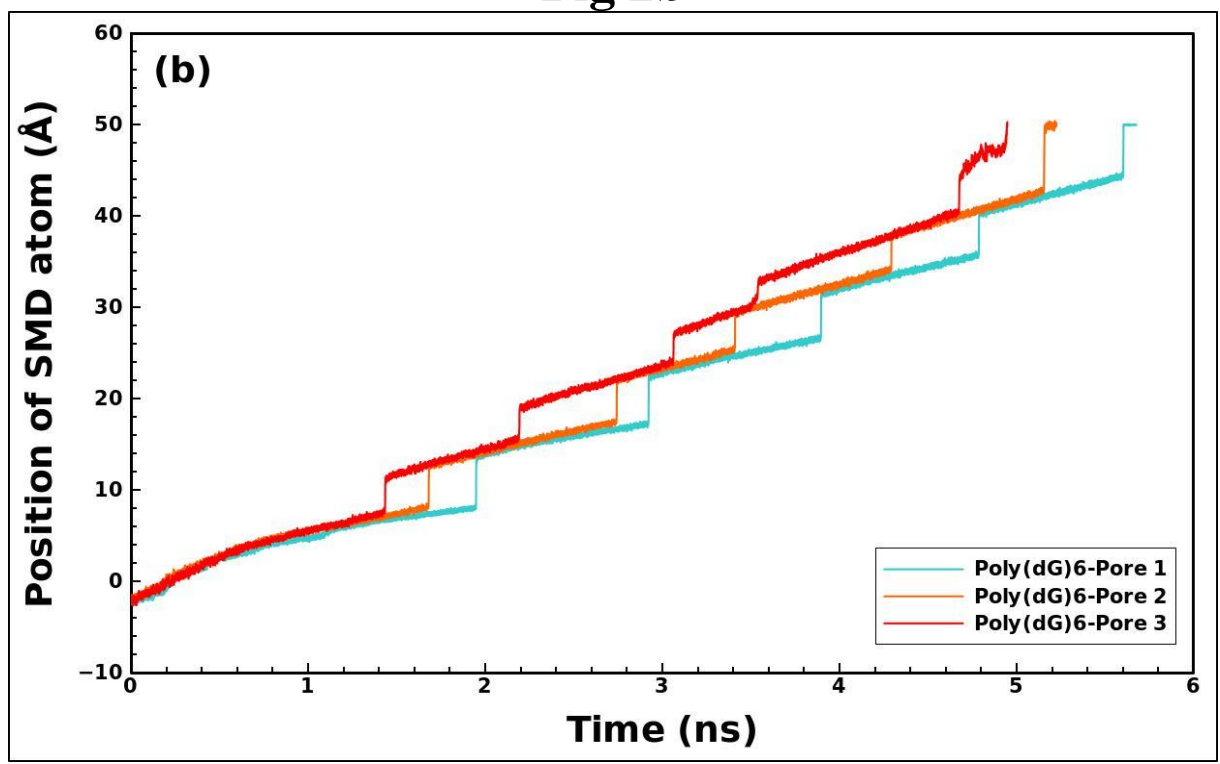

**Fig 2c**

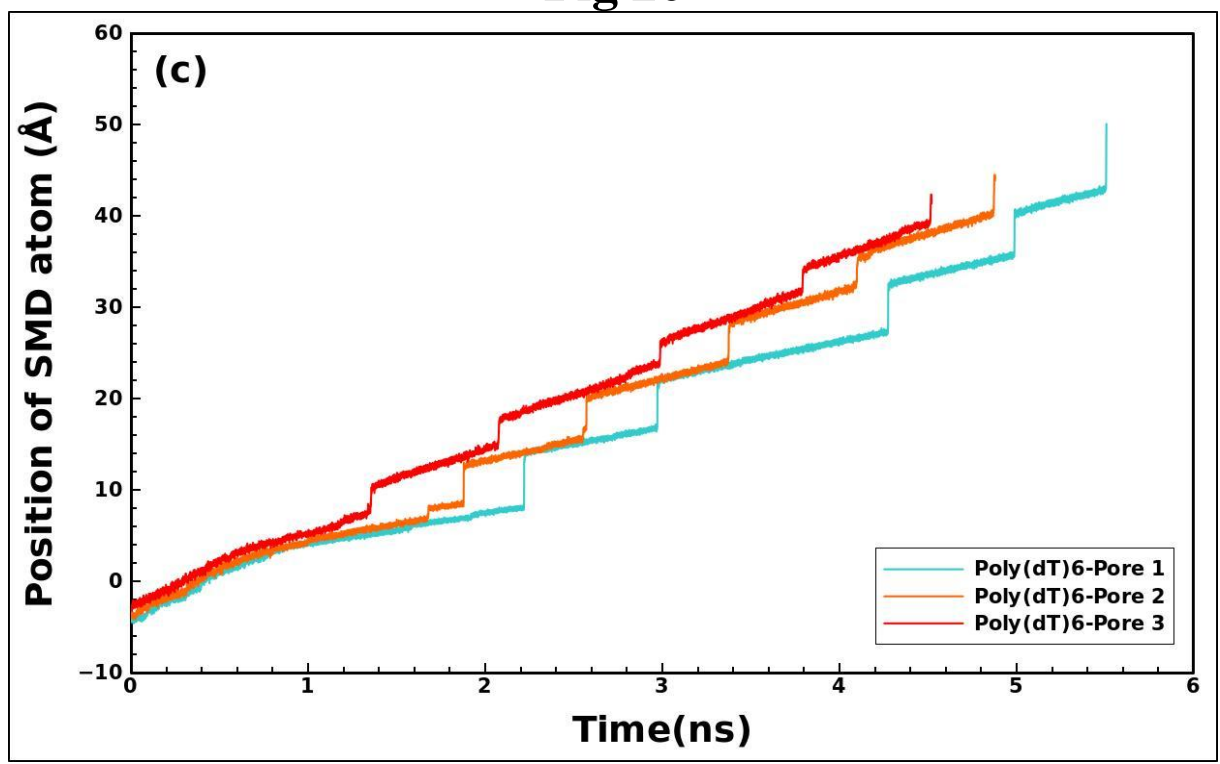

**Fig 2d**

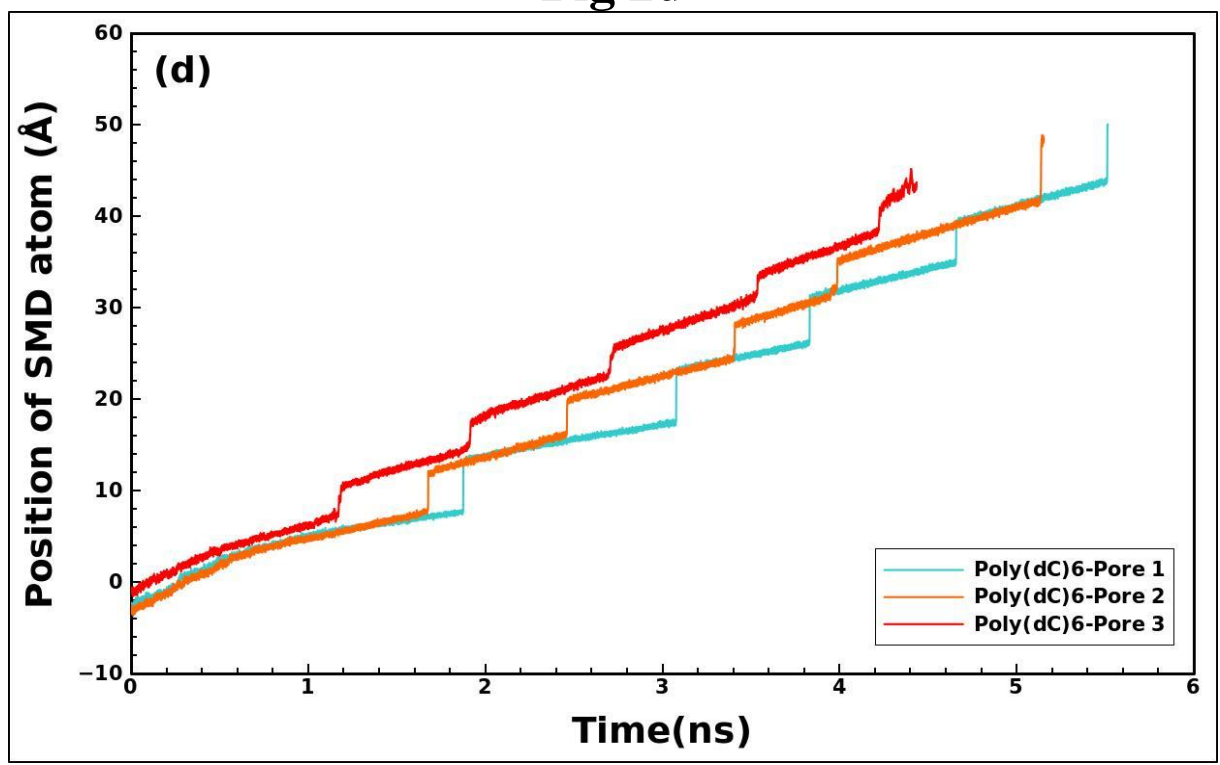

**Fig 2e**

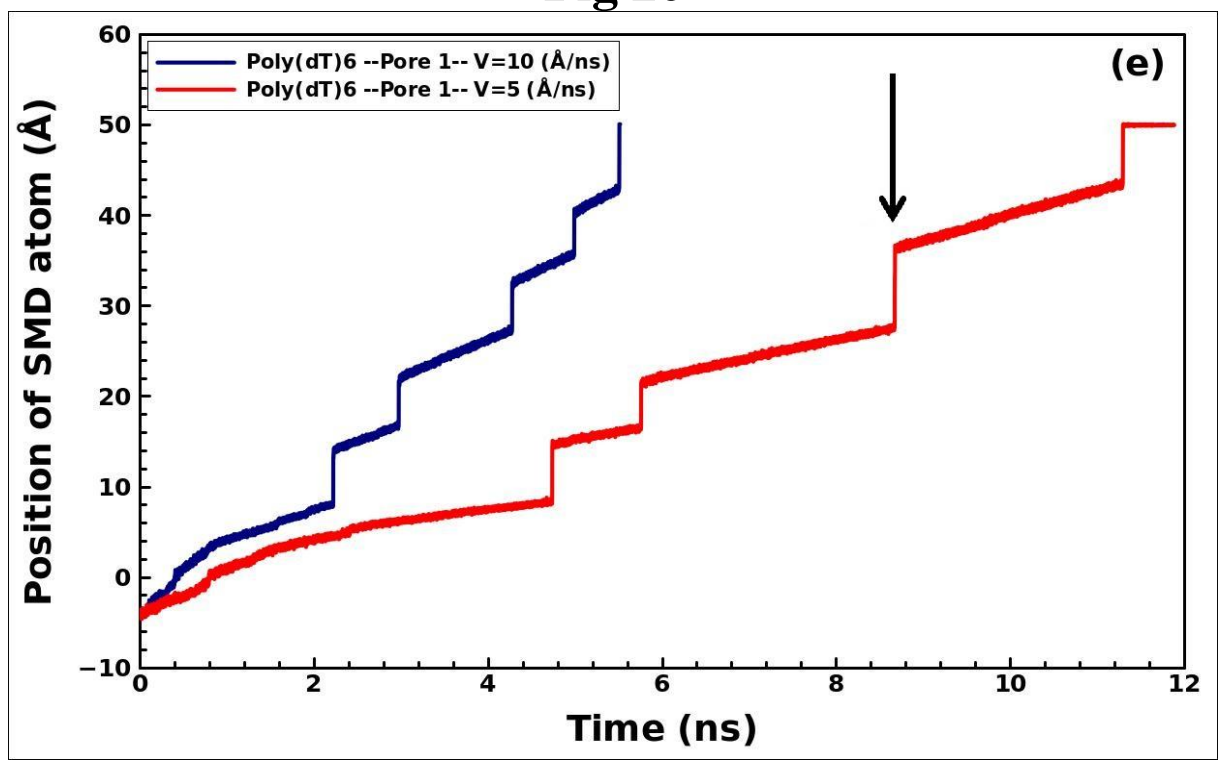

**Fig 2f**

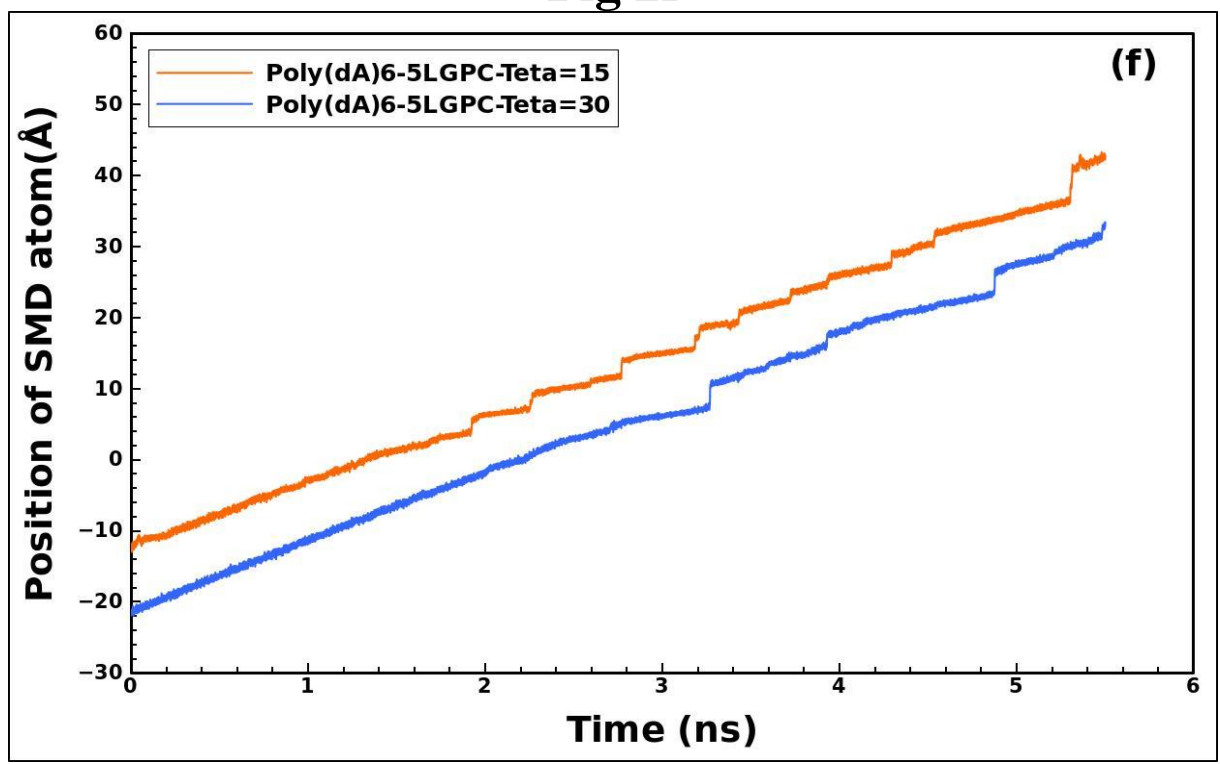

**Fig 3a**

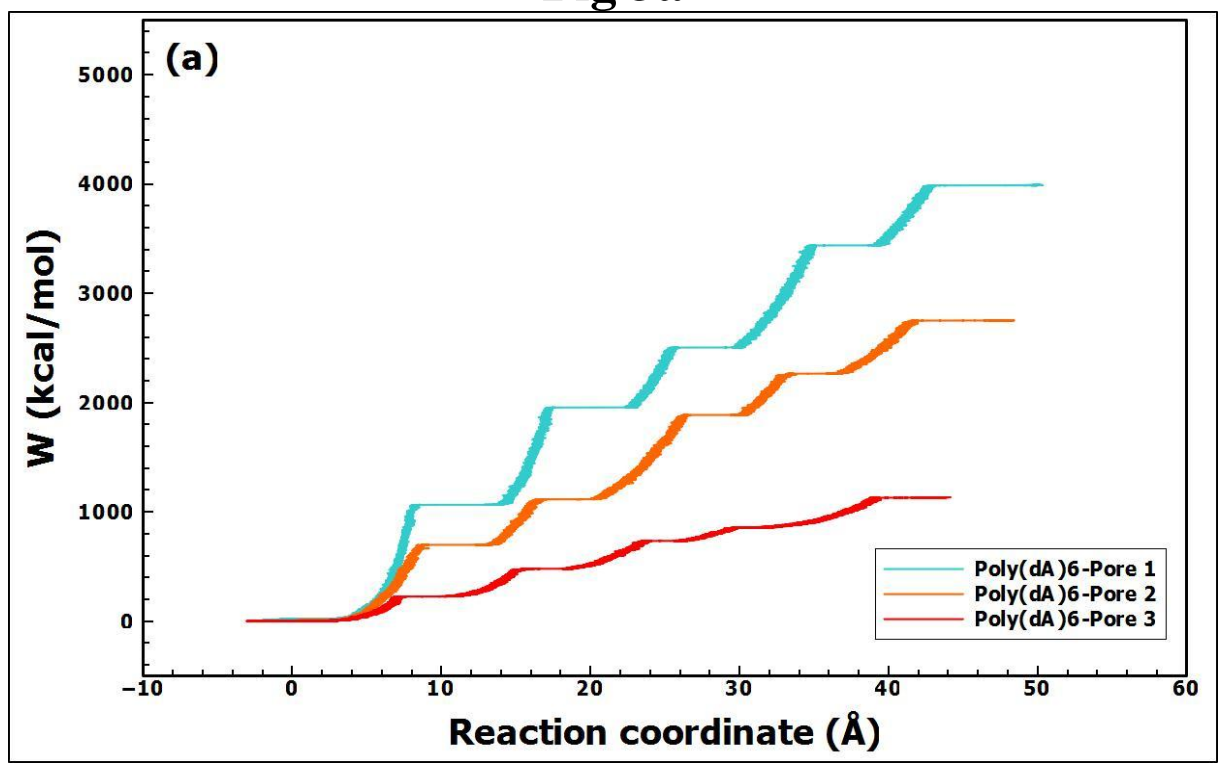

**Fig 3b**

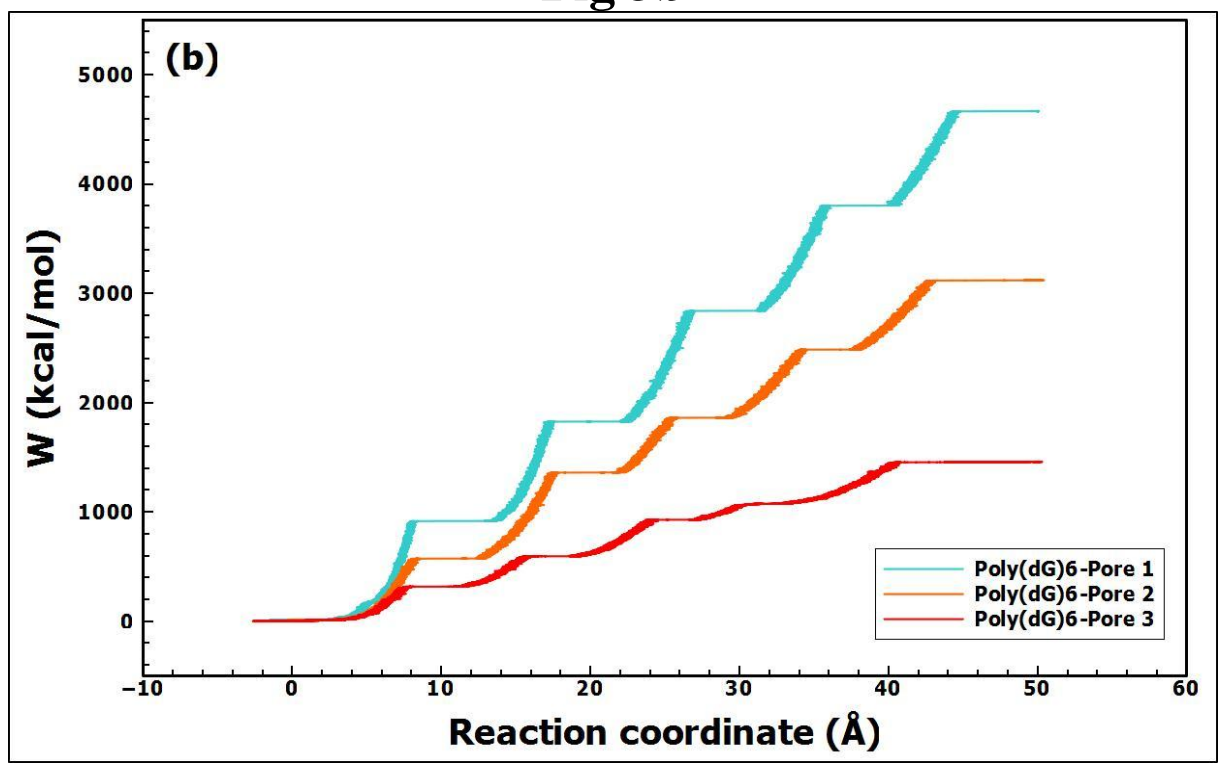

**Fig 3c**

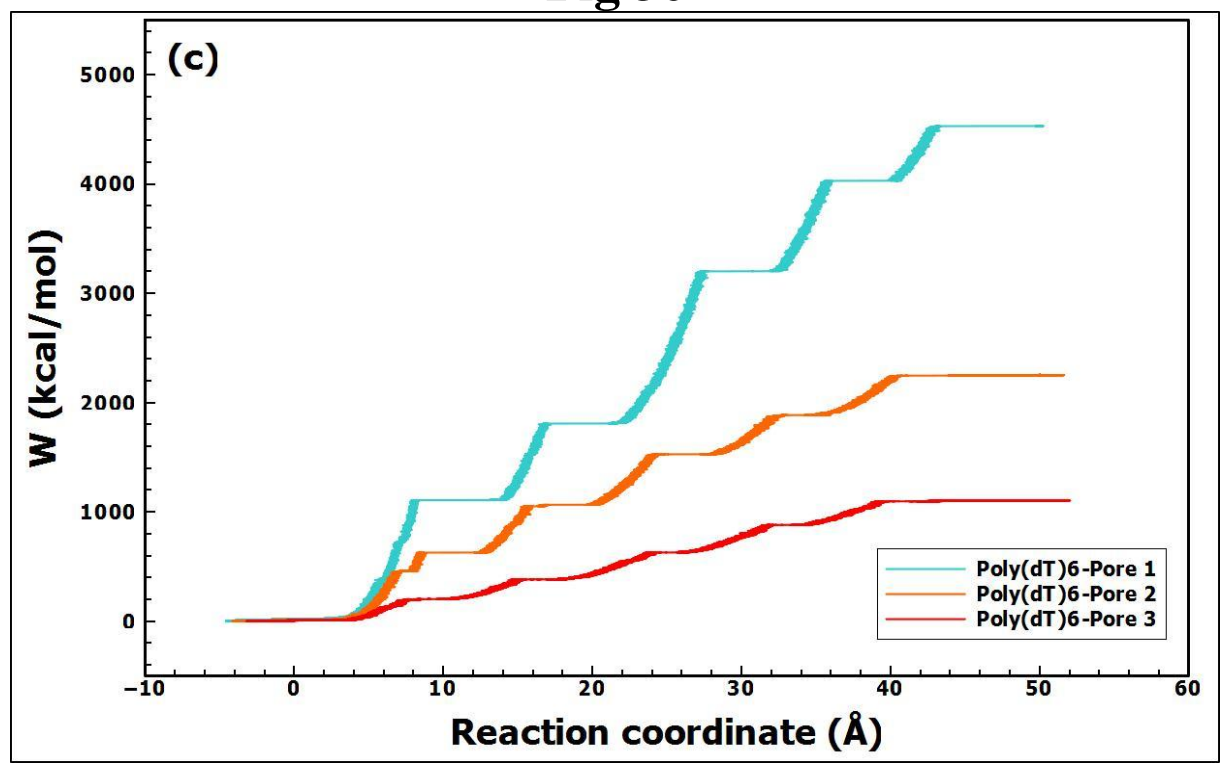

**Fig 3d**

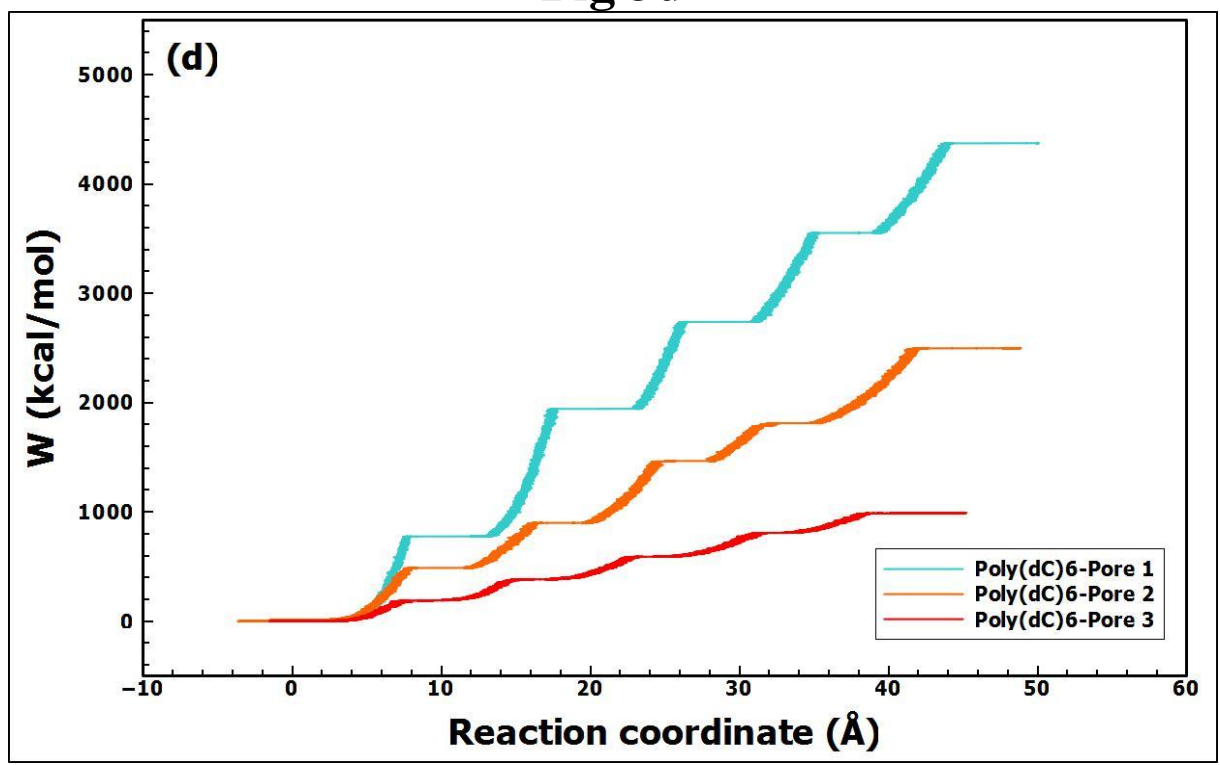

**Fig 3e**

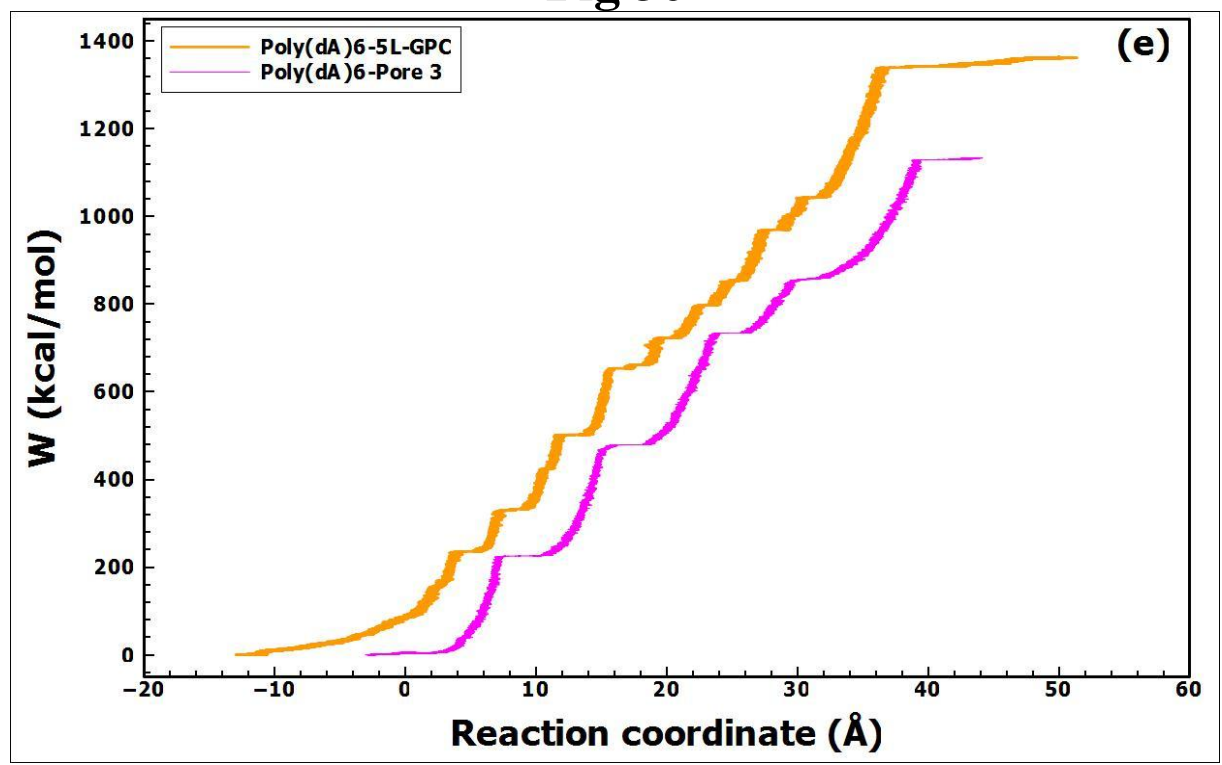

**Fig 4a**

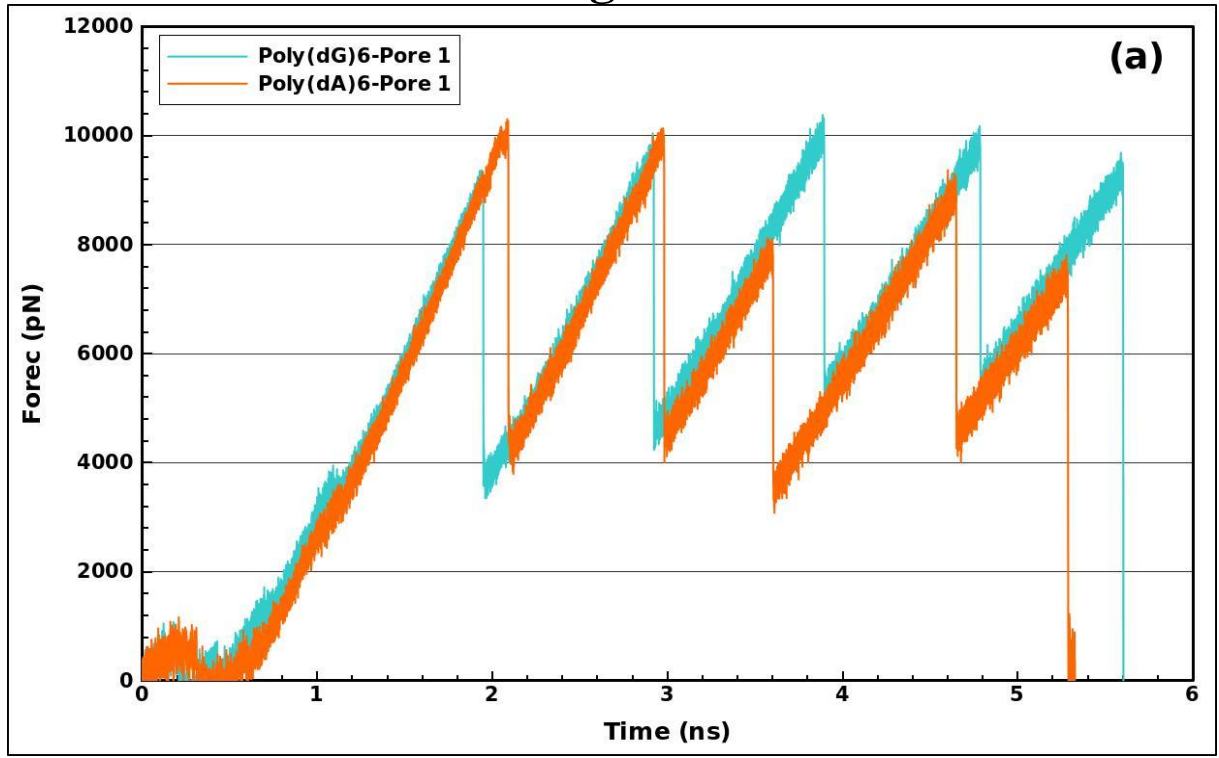

**Fig 4b**

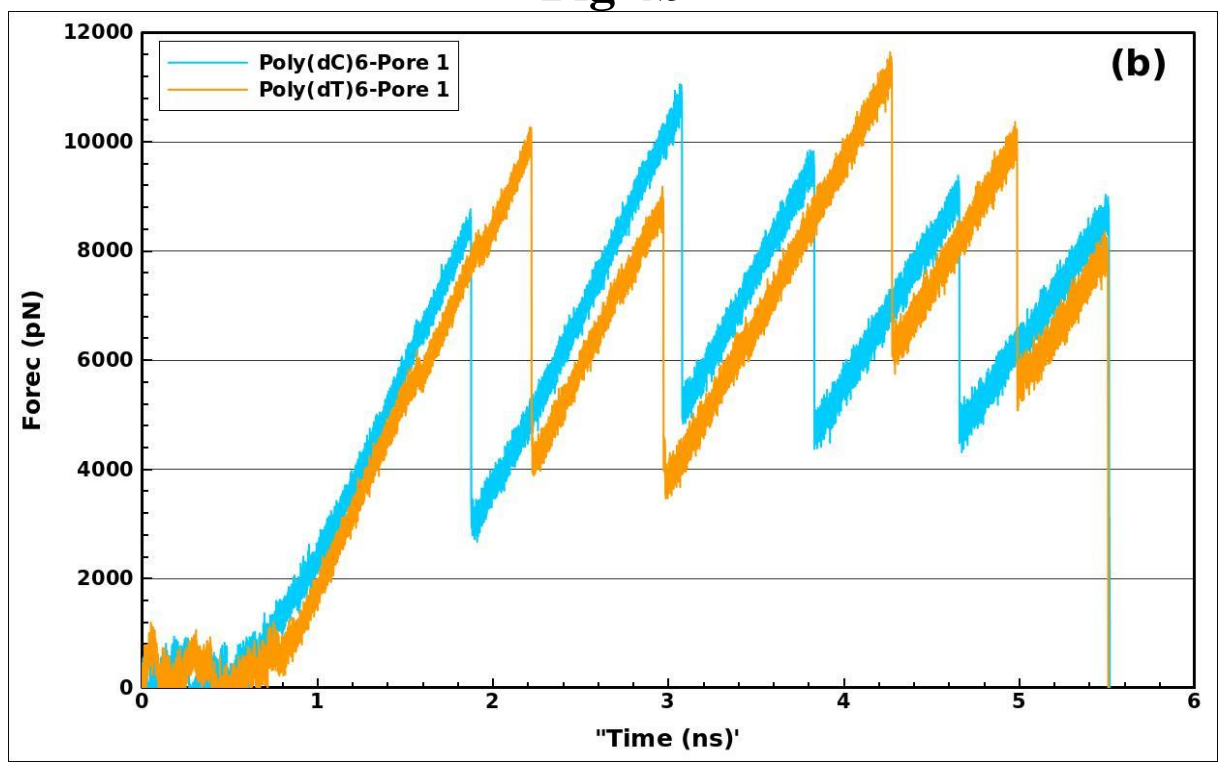

**Fig 4c**

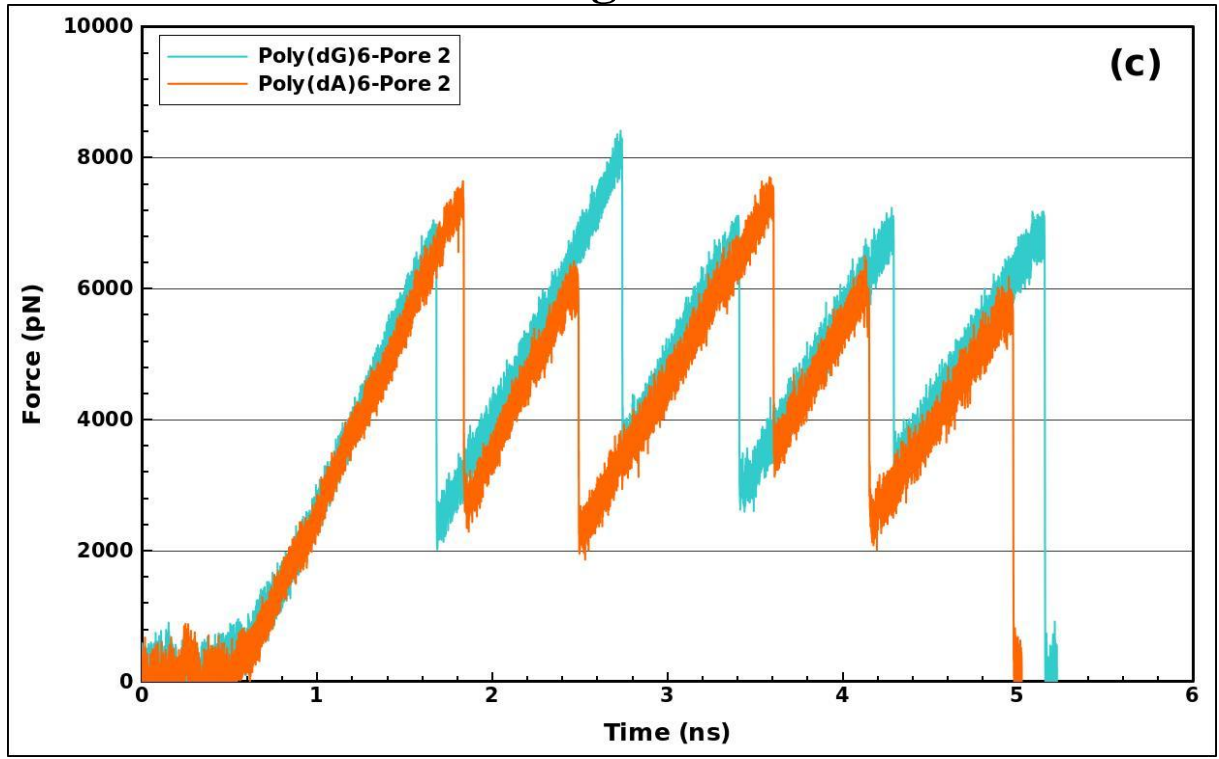

**Fig 4d**

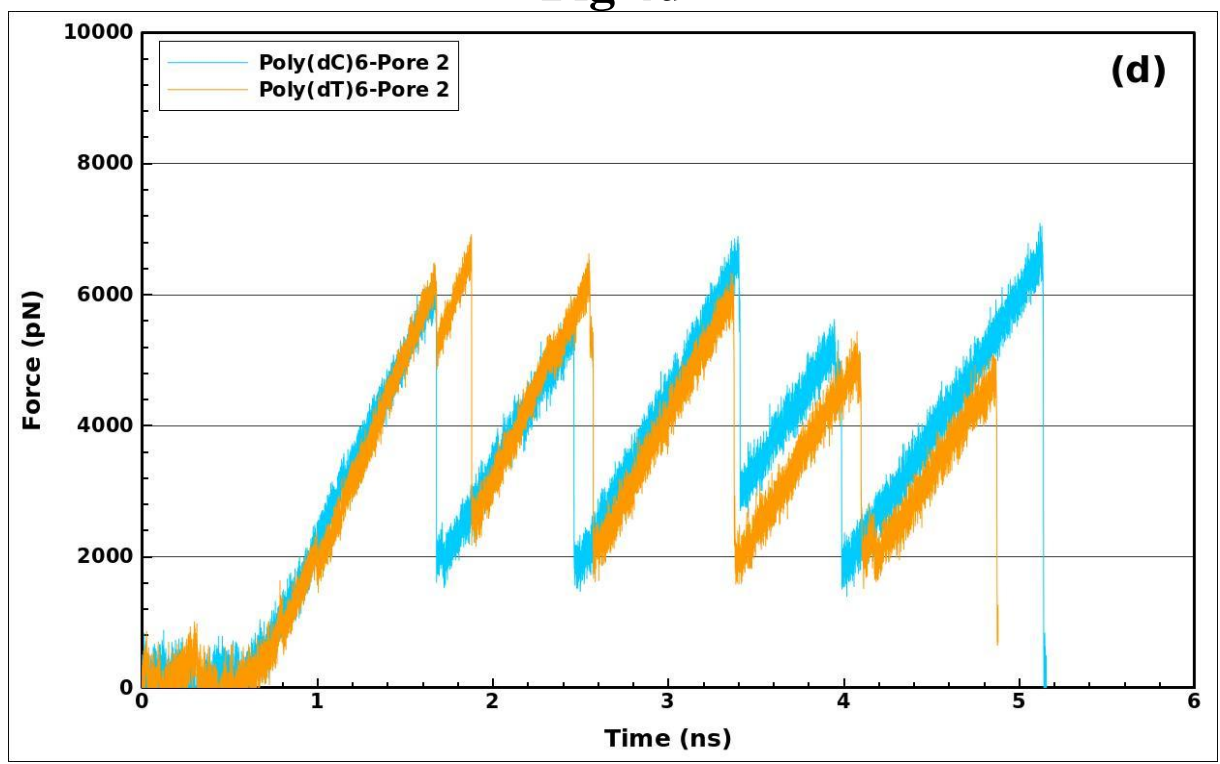

**Fig 4e**

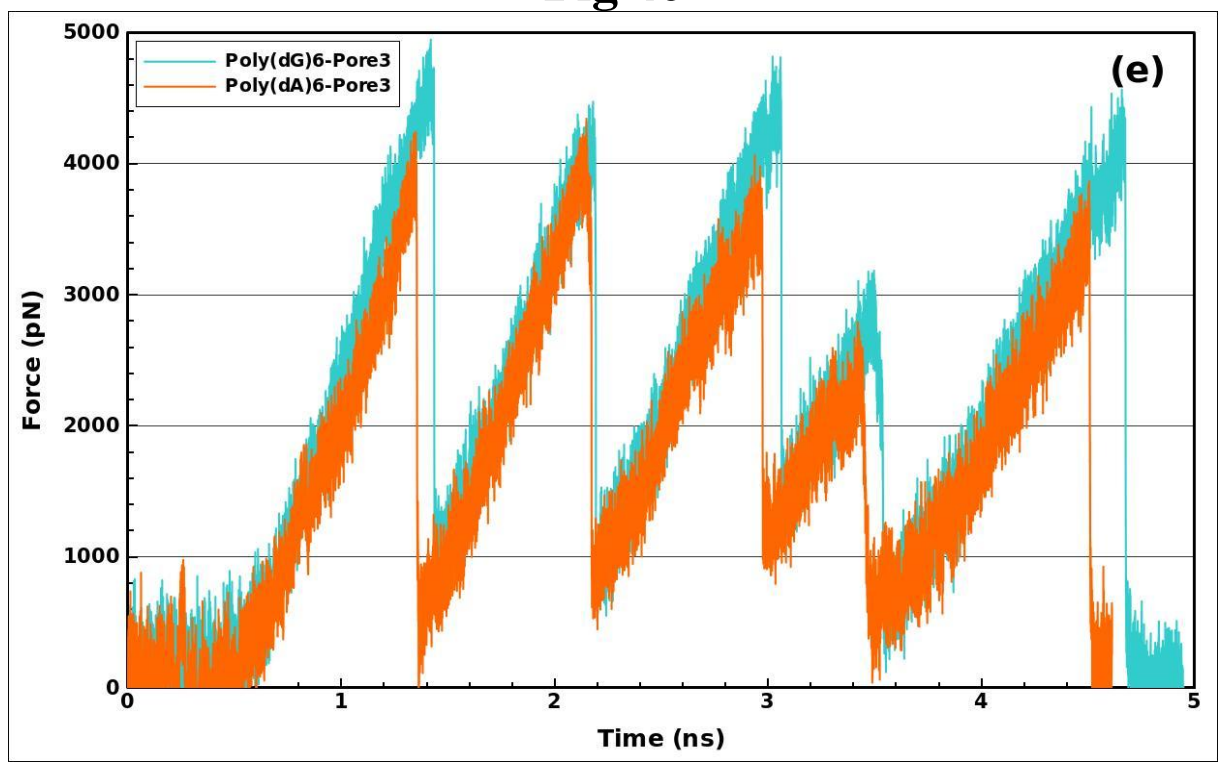

**Fig 4f**

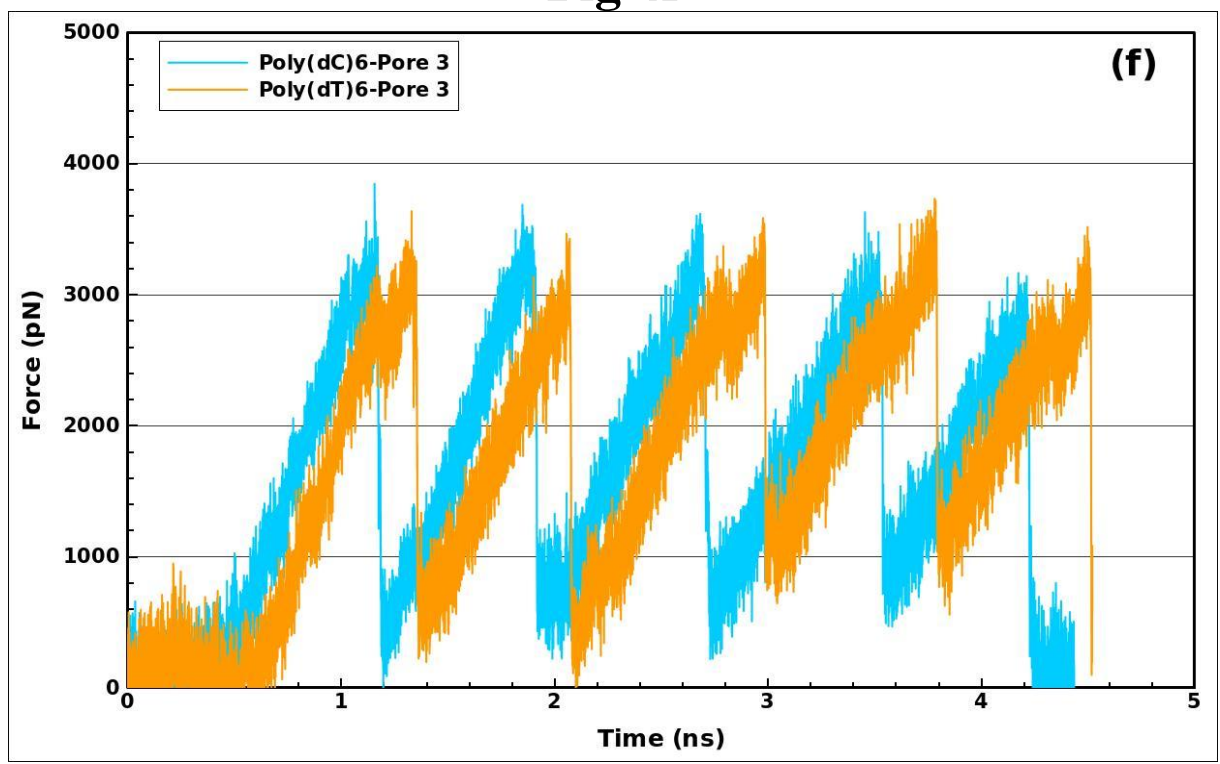

**Fig 4g**

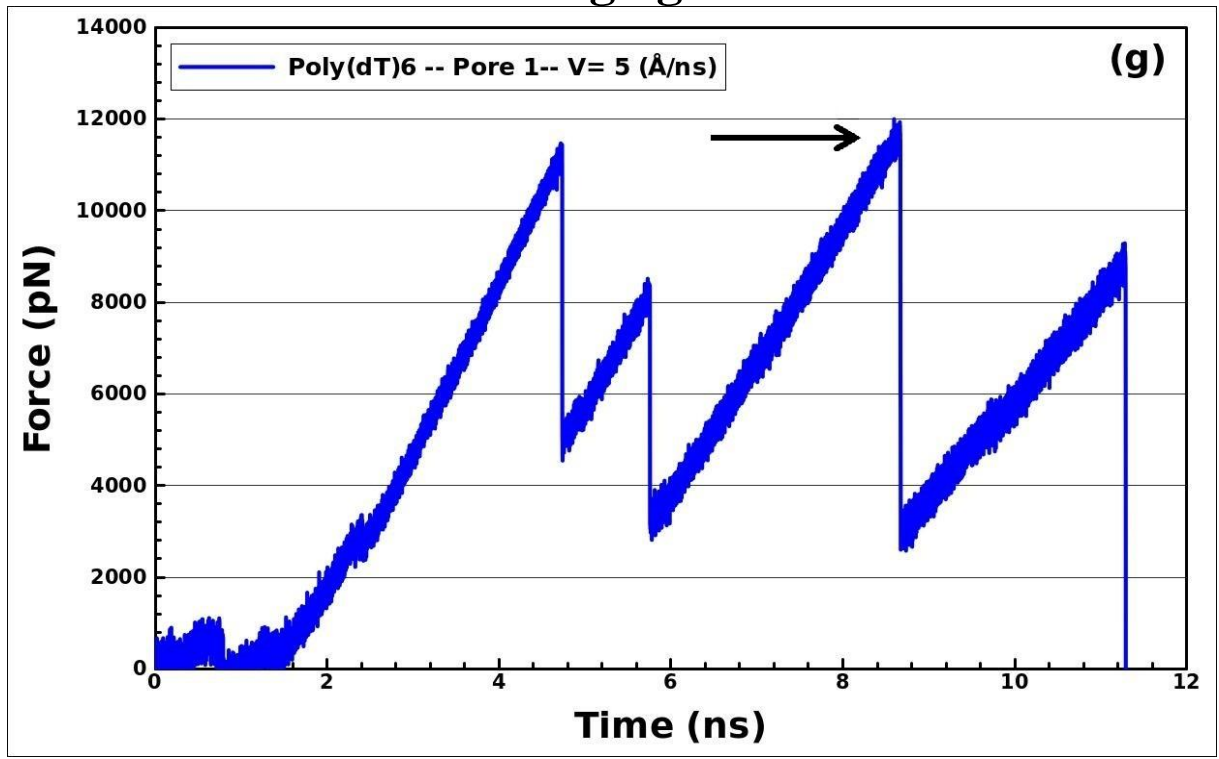

**Fig 4h**

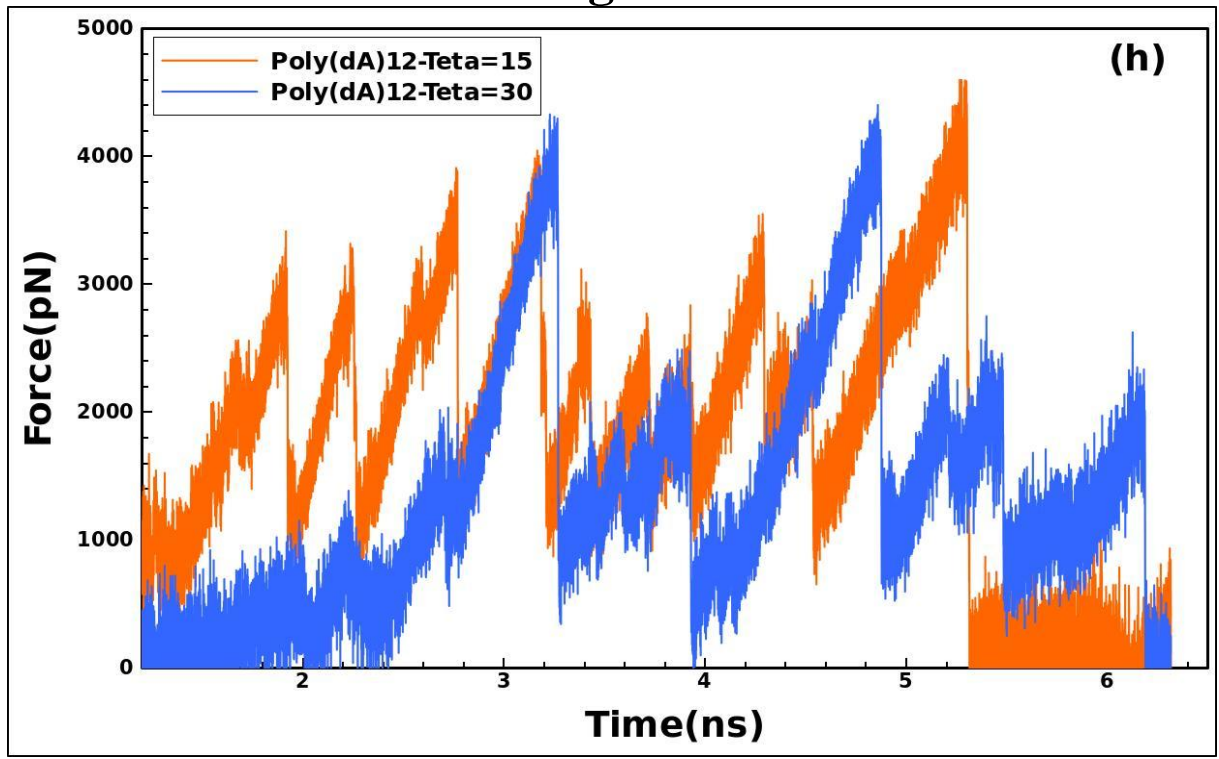

**Fig 5**

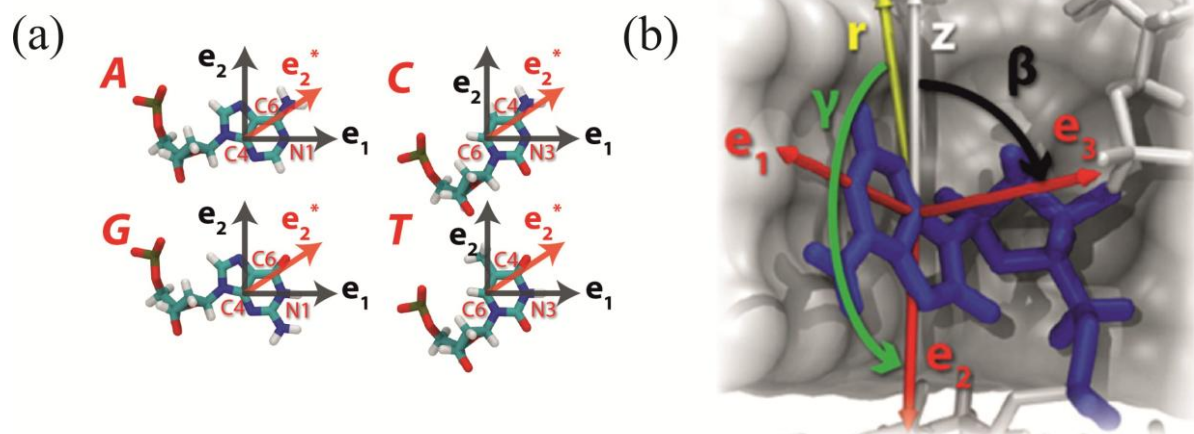

**Fig 6a**

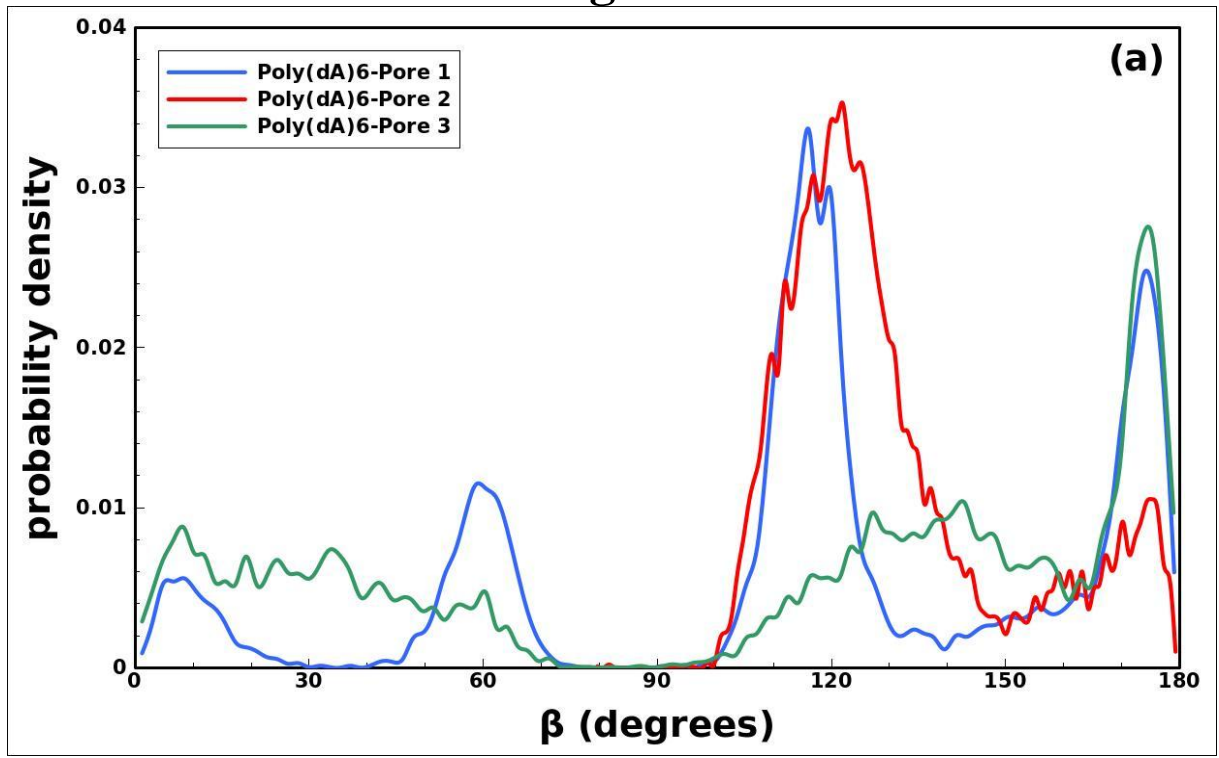

**Fig 6b**

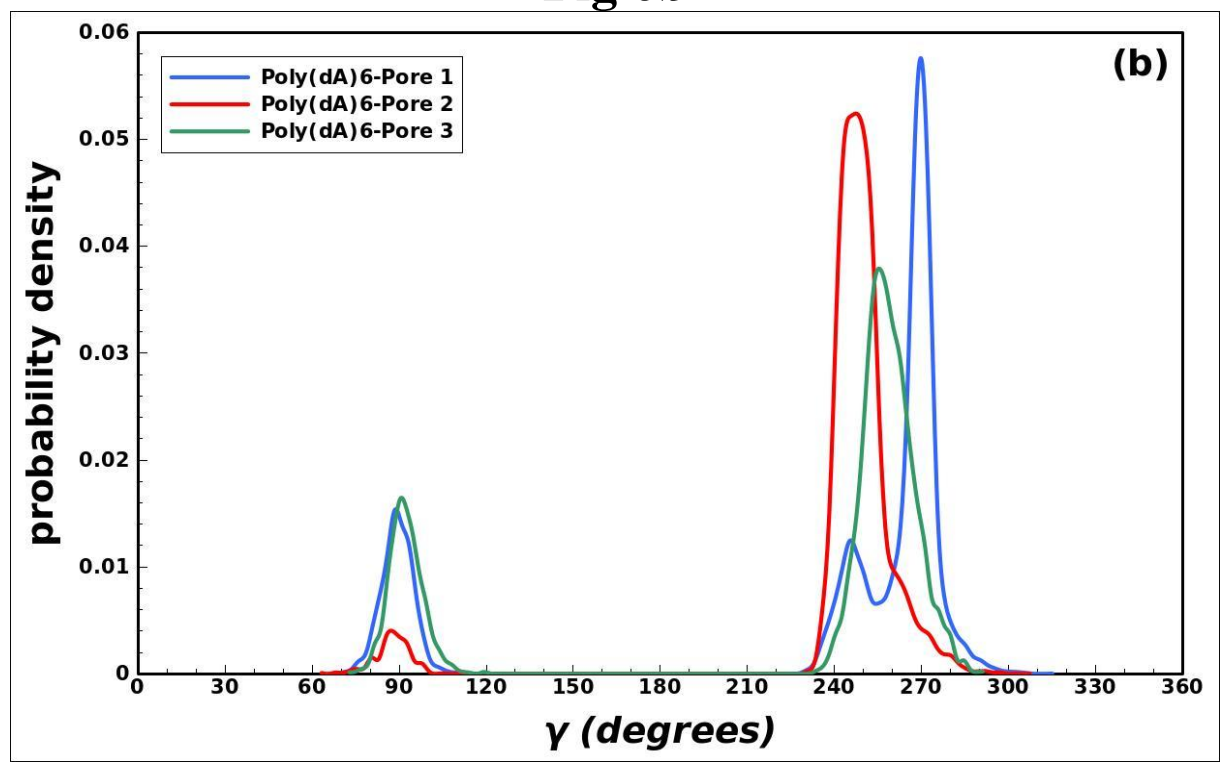

**Fig 6c**

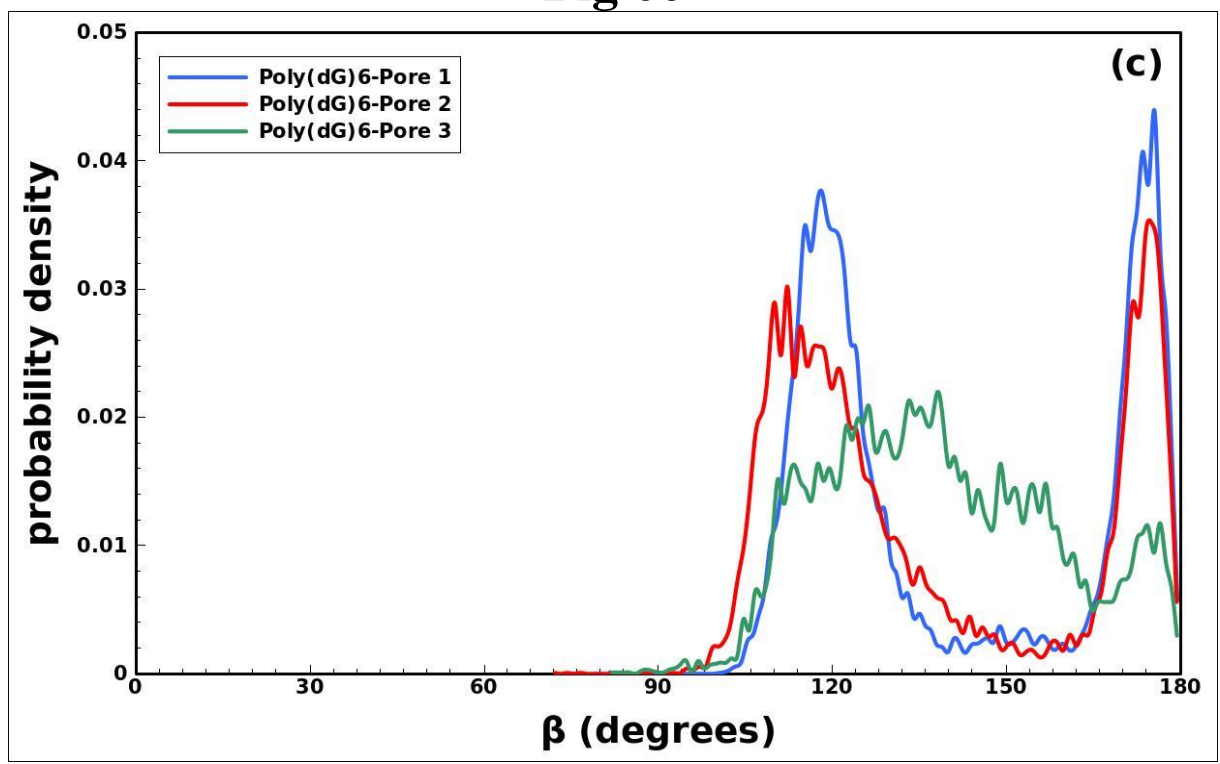

**Fig 6d**

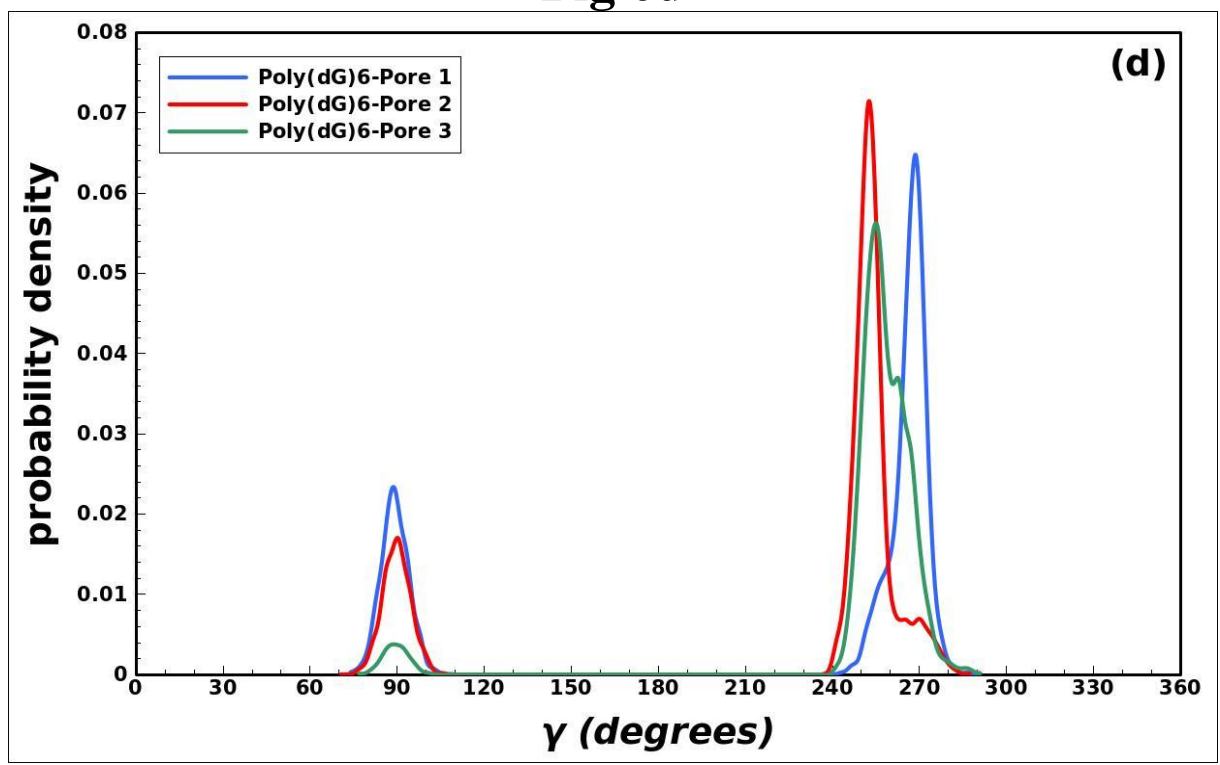

**Fig 6e**

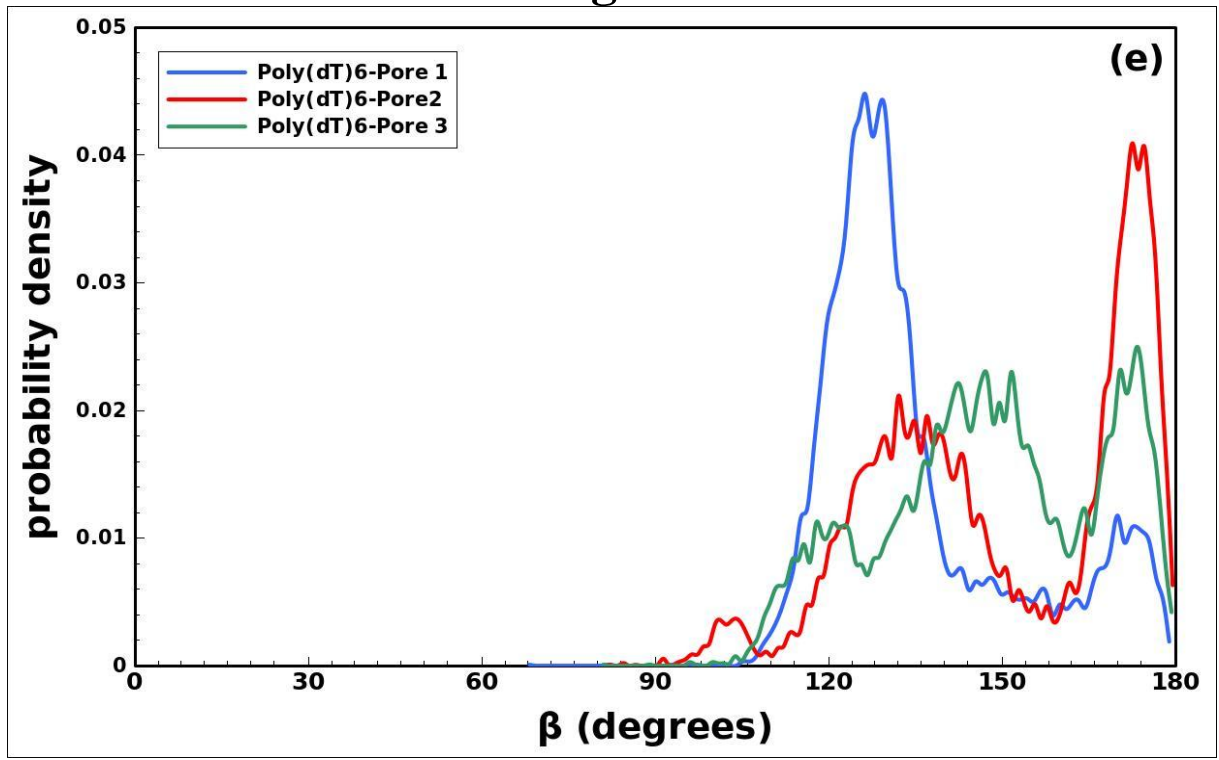

**Fig 6f**

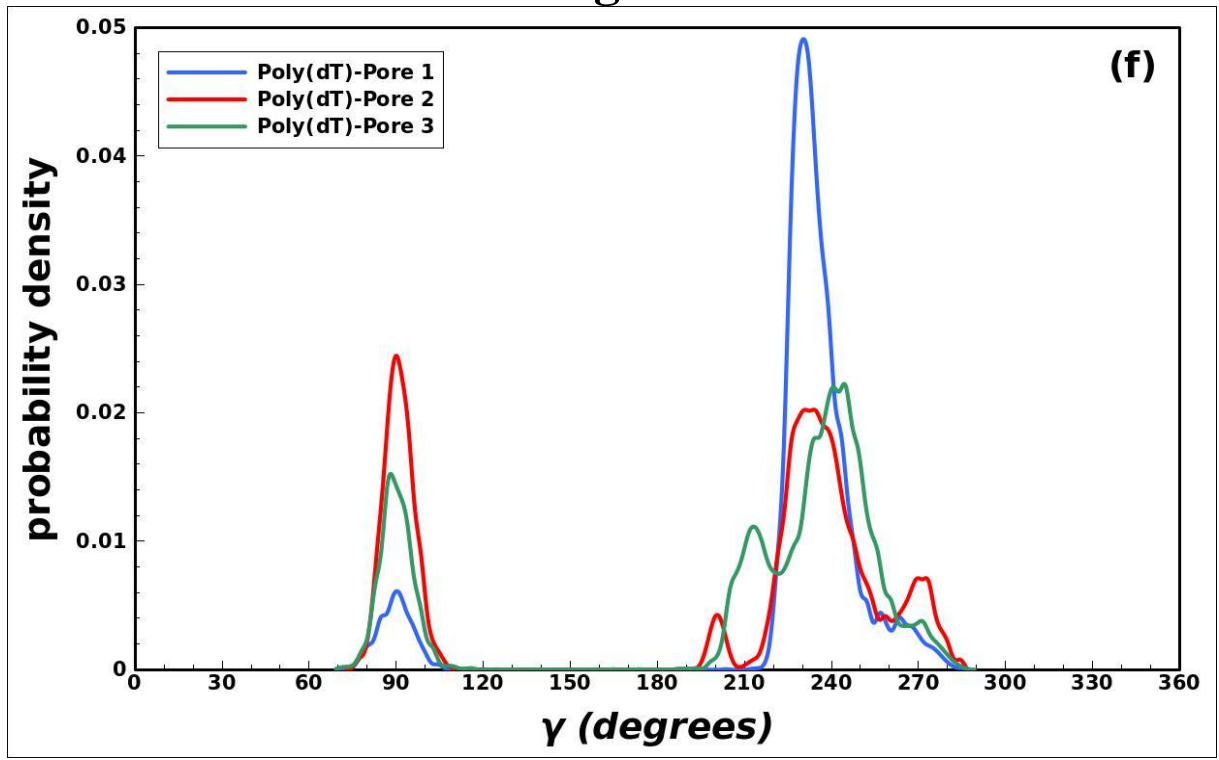

**Fig 6g**

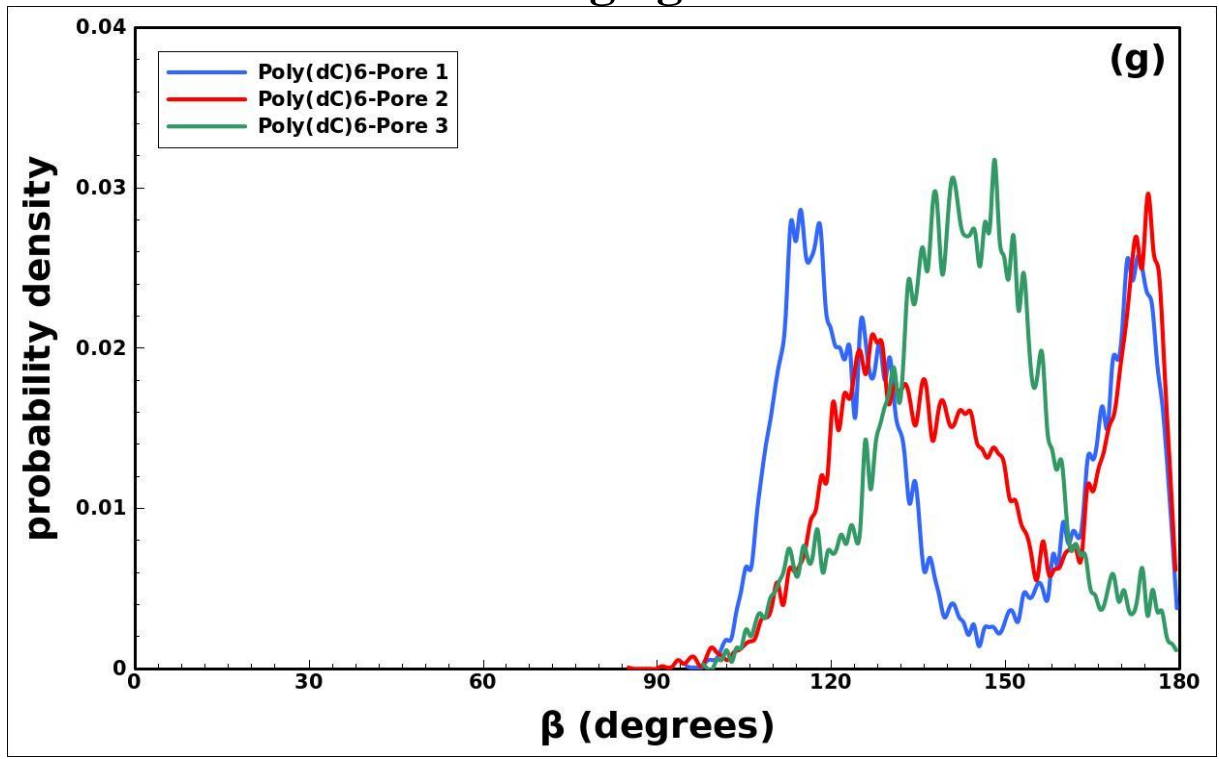

**Fig 6h**

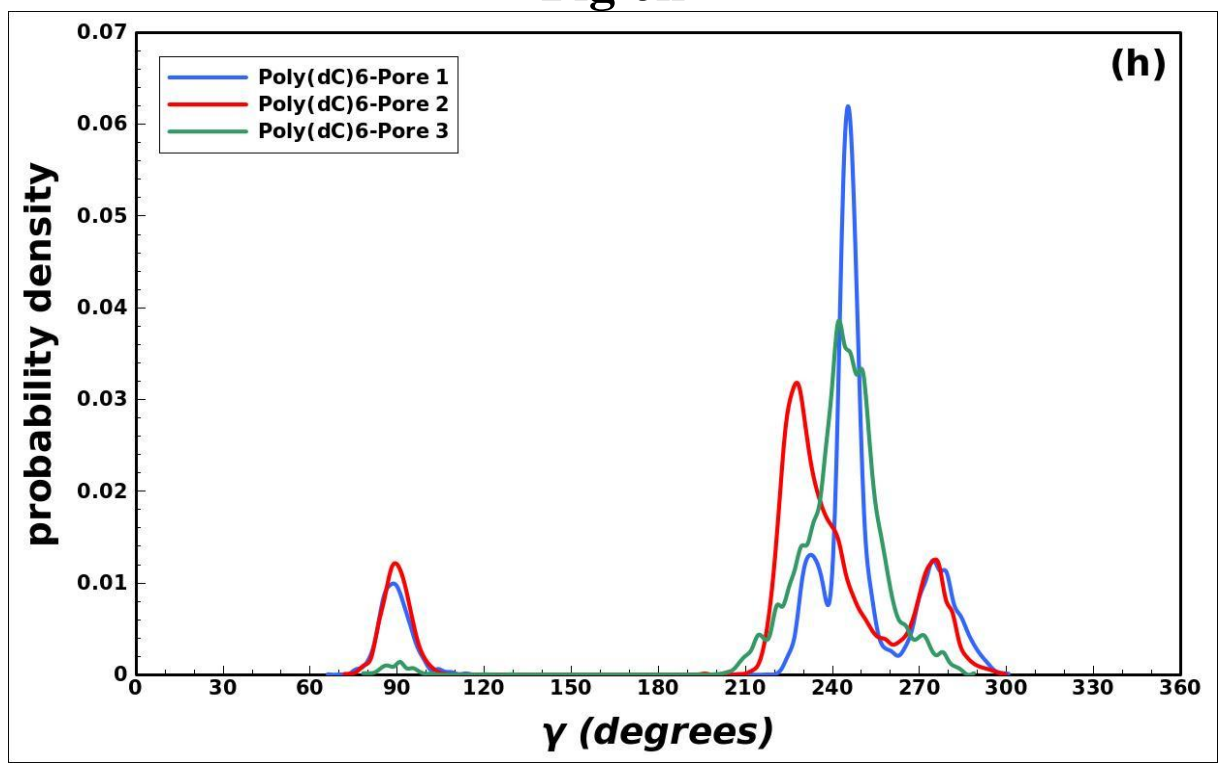